\documentclass[journal=jacsat,manuscript=article]{achemso}

\usepackage{chemformula} 
\usepackage[T1]{fontenc} 
\usepackage{ulem}

\author{Wen-Xin Jiang}
\affiliation{Department of Physics, State Key Laboratory of Quantum Functional Materials, and Guangdong 
Basic Research Center of Excellence for Quantum Science, Southern University of Science and Technology, Shenzhen, Guangdong 518055, China}

\author{Zhen-Hao Gong}
\affiliation{Department of Physics, State Key Laboratory of Quantum Functional Materials, and Guangdong 
Basic Research Center of Excellence for Quantum Science, Southern University of Science and Technology, Shenzhen, Guangdong 518055, China}

\author{Yuantao Chen}
\affiliation{Department of Physics, State Key Laboratory of Quantum Functional Materials, and Guangdong 
Basic Research Center of Excellence for Quantum Science, Southern University of Science and Technology, Shenzhen, Guangdong 518055, China}

\author{Zhigang Gui}
\email{Corresponding author: guizhigang@quantumsc.cn}
\affiliation{Quantum Science Center of Guangdong-Hong Kong-Macao Greater Bay Area (Guangdong), Shenzhen 518045, China}
\affiliation{Department of Physics, State Key Laboratory of Quantum Functional Materials, and Guangdong 
Basic Research Center of Excellence for Quantum Science, Southern University of Science and Technology, Shenzhen, Guangdong 518055, China}

\author{Li Huang}
\email{Corresponding author: huangl@sustech.edu.cn}
\affiliation{Department of Physics, State Key Laboratory of Quantum Functional Materials, and Guangdong 
Basic Research Center of Excellence for Quantum Science, Southern University of Science and Technology, Shenzhen, Guangdong 518055, China}
\affiliation{Quantum Science Center of Guangdong-Hong Kong-Macao Greater Bay Area (Guangdong), Shenzhen 518045, China}

\title{Sliding Engineering Spin-Valley-Layer Coupling and Altermagnetism in Bilayer Antiferromagnetic Honeycomb Lattices}

\abbreviations{IR,NMR,UV}

\begin{document}







\begin{abstract}
Valley polarization and altermagnetism are two emerging fundamental phenomena in condensed matter physics, offering unique opportunities for information encoding in novel energy-efficient devices. However, achieving electrical control over these properties in a single material remains a significant challenge. Here, we propose a general strategy to realize ferroelectric-valley (FE-valley) and FE-altermagnetic coupling in bilayer antiferromagnetic (AFM) honeycomb lattices based on an effective four-band spin-full \(k\cdot p\) model. Our proposal is validated in bilayer MnPTe$_3$ through first-principles calculations. A spontaneous out-of-plane electric polarization occurs in AB- and BA-stacked configurations, reversibly switchable via interlayer sliding. Remarkably, polarization reversal simultaneously inverts both layer-resolved valley polarization and altermagnetic spin splitting. This dual control enables tunable layer-spin-locked anomalous valley Hall effects and an unprecedented magnetoelectric response in 2D antiferromagnets. Our work provides a design principle for  electrically programmable valleytronic and spintronic functionalities of 2D AFM materials, bridging fundamental symmetry-breaking mechanisms and practical device applications. 

\end{abstract}

\section{Introduction}

The emerging fields of valleytronics and altermagnetism have opened new frontiers in condensed matter physics by exploiting valley and spin degrees of freedom as complementary information carriers \cite{XiaoDi-2007-valley-PRL, YaoWang-2008-Valley_optelectronics-PRB, XiaoDi-2012-valley_MoS2-PRL, Schaibley-2016-valleytronics_Review-NRM,Vitale-2018-valleytronics_Review-Small, Pacchioni-2020-valleytronics_comment-NRM,Smejkal-2022-alter1-PRX,Smejkal-2022-alter2-PRX,zhu-2024-alter_observation_nature,bai-2024-alter-AFM}. Valley polarization, characterized by the energy inquivalence between time-reversal (\(\mathcal{T}\))  linked \(K\) and  \(K'\) points in momentum space, was first predicted in graphene \cite{XiaoDi-2007-valley-PRL} and later realized in transition metal dichalcogenides \cite{mak-2014-MoS2-Science,lee-2016-bilayer_MoS2-Nature_nano}. Recent developments in altermagnets, which exhibit momentum-locked spin splitting despite vanishing net magnetization, have revealed unique transport phenomena distinct from conventional ferromagnetic (FM) and antiferromagnetic (AFM) paradigms \cite{Smejkal-2022-alter1-PRX,Smejkal-2022-alter2-PRX,zhu-2024-alter_observation_nature,bai-2024-alter-AFM}. In particular, valley polarization and altermagnetic spin textures near the Fermi level can encode binary information, offering unprecedented opportunities for ultra-dense, non-volatile memory technologies \cite{Wu-2024-valley_alter-nanoletter,guo-2024-valley_alter-2024,Feng-2025-FE_valley_alter_PT-PRB}. Therefore, it is highly desirable to have valleytronic and altermagnetic properties coexisting in a single material.


Efforts to engineer tunable valley polarization have largely focused on coupling valley states to ferroic orders (e.g.,  ferroelectricity or magnetism).\cite{DuanChungang-2022-FE_valley_coupling_review-APL,PeiQi-2019-FE_valley_AgBiP2Se6/CrI3-ACS_AMI,DuanChungang-2020-FE_valley_CuInP2S6/MnPSe3-JMCC,MiWenbo-2018-AFE_valley_BiXO3/BiIrO3-ACS_AMI,MaXikui-2020-FE_valley_VTe2/Ga2S3-JMCC,zhai2021using,lai2019two,DuanChungang-2017-FE_valley_GeSe-2DMaterials,XiaoChengcheng-2018-FE_valley_Bi-AdvancedFM,LiLei-2019-AdsorpFE_valley_XCl2(C2N)6-AdvancedFM,LiuXingen-2020-FE_valley_bi_VS2-PRL,MaYandong-2022-FE_valley_bi_FeCl2-npj_CM,ZhangXianmin-2023-FE_valley_bi_YI2-NanoLetter,Gui_2025}. For instance, FE-valley coupling enables nonvolatile switching of valley polarization, as demonstrated in sliding FE bilayers like T-FeCl$_2$ \cite{MaYandong-2022-FE_valley_bi_FeCl2-npj_CM} and YI$_2$.\cite{ZhangXianmin-2023-FE_valley_bi_YI2-NanoLetter} Meanwhile, advances in altermagnetism have identified bulk materials (e.g., RuO$_2$\cite{vsmejkal-2020-RuO2,Bai-2022-RuO2},MnTe\cite{Lee-2024-MnTe,Mazin-2023-MnTe,krempasky-2024-MnTe} and CrSb\cite{vsmejkal-2022-CrSb,Reimers-2024-CrSb,Yang-2025-CrSb,Liuchang-2024-CrSb}) with momentum-locked spin splittings, but extending these properties to 2D van der Waals (vdW) systems, where external stimuli like twisting or electric fields can dynamically tailor functionalities, has only recently been explored. A promising strategy involves breaking \(\mathcal{PT}\) symmetry in AFM homobilayers through interlayer stacking, which can induce altermagnetic behavior. \cite{Pan-2024-stacking_alter,Sun-2025-stacking_alter,Sun-2025-alter}. From the symmetry point of view, the coexistence of ferrovalley, altermagnetism and ferroelectricity are not exclusive. Such synergistic couplings would enable simultaneous control of spin, valley and charge degrees of freedom through a single external stimulus. Despite these strides, the simultaneous realization and electrical control of both valley polarization and altermagnetism in a single material system remains an outstanding challenge in the field\cite{2023umar,LiuQihang-2025-FEalter,DuanXuankai-2025-AFEalter,Sun-2025-alter,Wu-2024-valley_alter-nanoletter,Feng-2025-FE_valley_alter_PT-PRB}.

In this work, based on a four-band spin-full low-energy effective \(k\cdot p\) model, we design a system with coexisting FE-valley and FE-altermagnetic couplings formed by a 2D AFM honeycomb lattices. Our proposal is further demonstrated using bilayer MnPTe$_3$ as an example through first-principles calculations. A spontaneous out-of-plane electric polarization occurs in AB- and BA-stacked configurations, which is reversibly switchable via interlayer sliding. This polarization lifts the valley degeneracy, leading to pronounced dynamic valley splitting. Moreover, the polar stacking breaks the \(\mathcal{PT}\) symmetry inherent in the monolayer structure while keeping a vertical mirror symmetry connecting opposite magnetic sublattices, resulting in altermagnetism with substantial momentum-dependent spin splitting. Remarkably, Reversing the FE polarization inverts both the layer contribution, the associated valley and spin splittings near the Fermi level, thereby enabling tunable layer-spin-locked anomalous valley Hall transport and efficient magnetoelectric couplings. These findings establish 2D AFM materials as veratile platforms for next-generation devices that integrate valleytronics, altermagnetic spintronics, and ferroelectrics in a single 2D material, opening new avenues for multifunctional quantum devices.

\begin{figure*}
\includegraphics[scale=0.5]{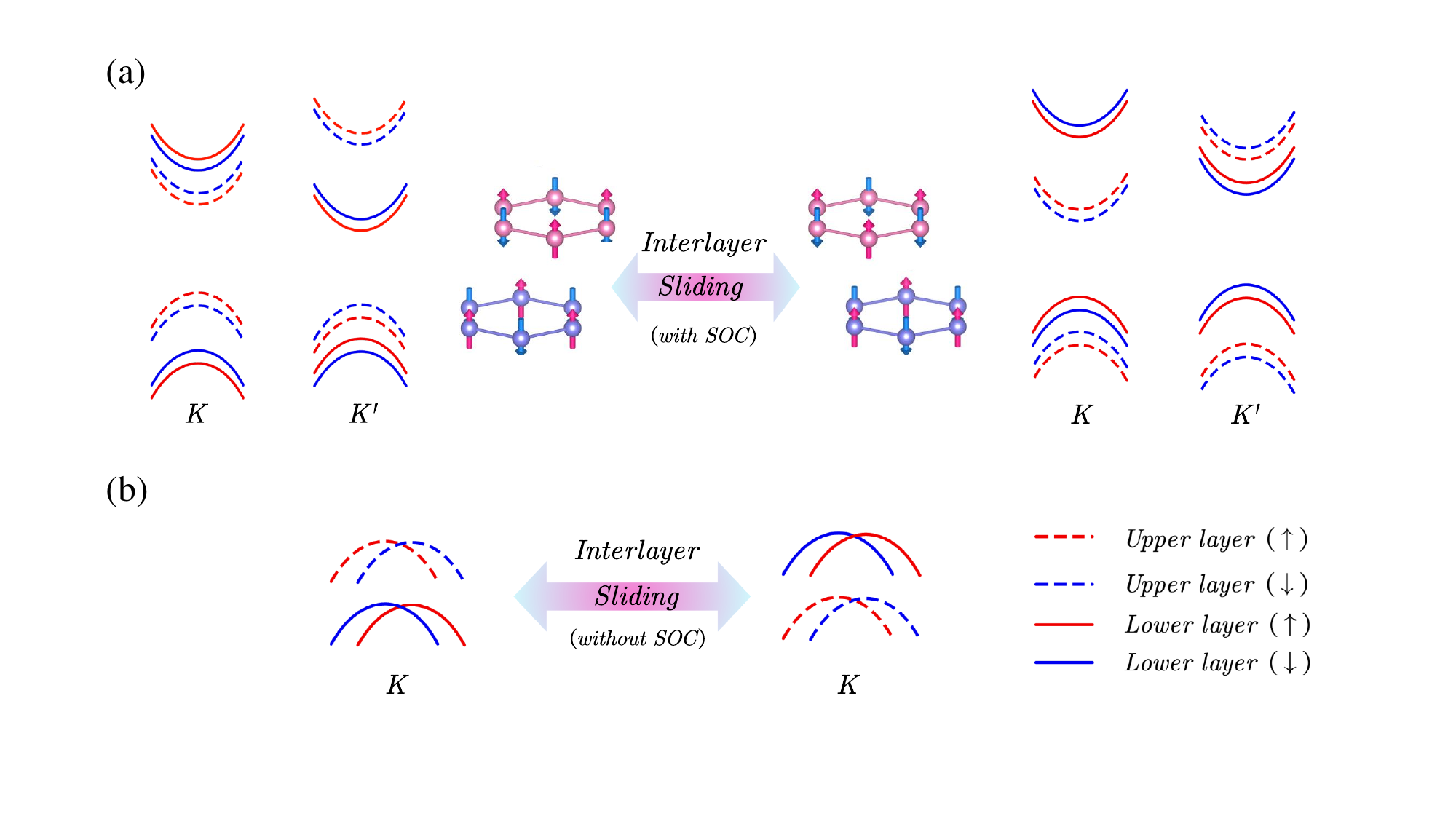}
\caption{Schematic illustration of layer–spin–valley coupling and altermagnetism in a bilayer  AFM honeycomb lattice. (a) Valley configurations with spin-orbit  coupling (SOC) at the \(K\) and \(K'\) points for AB (left) and BA (right) stacking. Solid/dotted lines represent upper/lower layers; red/blue indicates opposite spin. Reversible interlayer sliding between AB and BA stacking switches the FE polarization  (P↓ $\leftrightarrow$ P↑), simultaneously inverting the layer-resolved valley polarization. (b) Momentum-dependent spin splitting around the \(K\) valley (without SOC), demonstrating electrically switchable altermagnetic behavior with layer-locked spin polarization via interlayer sliding.}\label{stacking}
\end{figure*}

\section{Results and discussion}

\subsection{A. Four-band k·p model for bilayer  AFM honeycomb lattice}

To describe the FE-valley and FE-altermagnetic coupling system formed by a bilayer intralayer AFM honeycomb lattice, we develop a minimum four-band spin-full low-energy \(k\cdot p\) model. This effective Hamiltonian captures the essential interplay between charge, spin, valley, and layer degrees of freedom while elucidating their distinct roles in shaping the electronic structures near the Fermi level. Without loss of generality, the constituting monolayer should possess band extrema at the high-symmetry $K$ points and at least one symmetry operation connecting the two magnetic sublattices (for instance, a vertical mirror, which is compatible with the out-of-plane ferroelectricity in polar stacking bilayers). The resulting bilayer Hamiltonian takes the following form:
\[
H_k = 
\begin{bmatrix}
    H_k^u & H_\perp \\
    H_\perp & H_k^l
\end{bmatrix},
\]
where \(H_\perp\) represents the interlayer hopping term, which is typically negligible due to weak vdW interactions. The layer-resolved Hamiltonians \(H_k^{u}\) and \(H_k^{l}\) describe the upper and lower layers, respectively, with the lower-layer Hamiltonian \(H_k^{l}\) expressed as:  

\begin{multline}
H_k^{l} = v_f\,(-\sigma_x\,\tau_z\,s_0\,k_x + \sigma_y\,\tau_0\,s_0\,k_y) +m \, \sigma_z \,\tau_0 \, s_z  + \delta\,\sigma_0\,\tau_0\,s_{z}\,k_y - \lambda\,\sigma_z\,\tau_z\,s_z-\phi \,\sigma_z\,\tau_z\, s_0 + \sigma_0\,\tau_0\,s_{0}\,U_E,
\end{multline}
where the first term adopts the same low-energy effective \(k\cdot p\) model for graphene (keeping linear terms only), the second term is the mass term responsible for gap opening\cite{FengJi-2013-MnPX3-Valley-PNAS}, the third term is the spin splitting (keeping the lowest order) without spin-orbit coupling (SOC) (see supporting information for more details), the fourth and fifth terms mimic the spin-valley coupling from the simultaneous breaking of time reversal and spatial inversion symmetry (\(\mathcal{P}\)) originated from SOC \cite{WeiWei-2023-Mn2P2X3Y3-APL}, and the last term accounts for half of the electrostatic potential difference between the two layers due to the out-of-plane dipoles present in the polar  stacking configurations. Here, \(v_f\) is the Fermi velocity, \(\mathbf{k} = (k_x, k_y)\) the crystal momentum, and \(s_i\), \(\tau_i\), \(\sigma_i\) (\(i = 0, x, y, z\)) are Pauli matrices operating in spin, valley, and isospin degrees of freedom, respectively. \(\lambda\) and \(\phi\) are constant matrices. \(\delta\) and \(U_E\) are constants. The upper layer Hamiltonian \(H_k^{u}\) is obtained by applying a 
\(C_{2z}\) rotation to \(H_k^{l}\), which exchanges the \(K\) and \(K'\) valleys and reverses the sign of all terms containing \(\tau_z\), yielding:
\begin{multline}
H_k^{u} = v_f\,(-\sigma_x\,\tau_z\,s_0\,k_x + \sigma_y\,\tau_0\,s_0\,k_y) +m \, \sigma_z \,\tau_0 \, s_z -\delta\,\sigma_0\,\tau_0\,s_{z}\,k_y - \lambda^{'}\,\sigma_z\,\tau_z\,s_z-\phi^{'}\,\sigma_z\,\tau_z\, s_0 - \sigma_0\,\tau_0\,s_{0}\,U_E,
\end{multline}
Meanwhile, the out-of-plane electric dipole in the bilayer system produces an asymmetric potential landscape, inducing 
distinct electrostatic environments for the upper and lower layers. This potential difference manifests as layer-dependent 
coupling constants (\(\lambda^{'}\) and \(\phi^{'}\)), as demonstrated in previous study\cite{2007Igor}.. 

The total Hamiltonian reveals a nature mechanism for nonvolatile control of the layer-resolved valley polarization. As shown in Fig. 1(a), the relative energy shifts from the fourth and fifth terms in Eqs. (1) and (2) produce distinct low-energy band dispersions around the \(K\) and \(K'\) valleys for opposite electric polarization directions. Under downward polarization, the conduction band minimum (CBM) occurs at \(K\) valley with dominant upper-layer contribution, while the lower-layer-derived CBM at \(K'\) appears at slightly higher energy. Taking  
\(\lambda\), \(\phi\), \(\lambda^{'}\) and \(\phi^{'}\) as positive constants in Eqs. (1) and (2), the spontaneous 
valley polarization can be quantified as \(-\lambda-\phi + \lambda^{'}+\phi^{'} + 2\, U_E\).

Remarkably, reversing the electric polarization direction maintains the same magnitude of valley polarization while inverting the layer-specific band alignment: the CBM shifts to the \(K'\) valley (lower-layer dominated) while upper-layer-derived 
\(K\) valley state rises slightly in energy (Fig. 1(a)). This demonstrates robust nonvolatile switching of layer-resolved valley 
polarization. An analogous effect also occurs in the valence band for smaller SOC, where polarization reversal similarly exchanges  the layer contributions to the valence band extrema (VBM).  
Notably, in the strong SOC regime (\(\lambda+\phi - \lambda^{'}-\phi^{'} > 2\, U_E\)), the system exhibits different behavior: while valley polarization persists, the band extrema remain locked to specific valleys regardless of polarization direction (Fig. S5). This valley-layer locking thus preserves electrically tunable valley splitting while preventing layer-selective control. 

Even in the absence of SOC, the Hamiltonian model demonstrate nonvolatile electrical control of spin textures. As shown in Fig. 1(b), the spin splitting around \(K\) valley can be reversibly switched simply by flipping the electric polarization direction, without requiring any alteration of the intrinsic magnetic order. This unique magnetoelectric response provides an energy-efficient pathway for all-electrical spin manipulation in AFM systems.

\subsection{B. Realization in real materials}

The proposed design principle is further validated in realistic materials via first-principles calculations. We select bilayer MnPTe$_3$ as a representative candidate, featuring intralayer AFM order and stacking-induced ferroelectricity. 
Monolayer MnPTe$_3$ possesses \(\mathcal{P}\) symmetry with a point group of D$_{3d}$ (Fig. S1) and adopts a Néel-type AFM ground state, which breaks local \(\mathcal{P}\) symmetry at the centers of Mn hexagons and leads to a substantial valley splitting of 103.6 meV at \(K\) and \(K'\) points. However, its  preserved \(\mathcal{PT}\) symmetry maintains spin degeneracy. To achieve nonvolatile control of both layer-resolved valley polarization and spin splitting from the breaking of \(\mathcal{PT}\) symmetry, we engineer polar bilayers via stacking operation. The polar stacking homobilayer (B) can be constructed in the following relation with respect to the corresponding single layer (S), where \(B = S + \hat{O}S\), with \(\hat{O} = \{O | t_0\}\) being a stacking operator consisting of rotational \(O\) and translational components \(t_0\). As shown in Fig. 2(a), two stable stable configurations emerge: AB stacking (\(\hat{O} = \{2_{001} | [\frac{1}{3}, 0, 0]\}\)), and BA stacking (\(\hat{O} = \{2_{001} | [\frac{2}{3}, 0, 0]\}\)). The AB (BA) stacking involves a lateral shift of \([ \frac{1}{3}, 0, 0 ]\) (\([ \frac{2}{3}, 0, 0 ]\)). Due to the broken in-plane mirror symmetry, a net downward electric polarization of 0.33 pC/m develops in the AB stacking, while an equal upward polarization is present in the BA stacking. These degenerate FE states are seperated by an energy barrier of 71.5 meV (Fig. 2(b)), enabling reversible switching via interlayer sliding. 


\begin{figure*}
\includegraphics[scale=1]{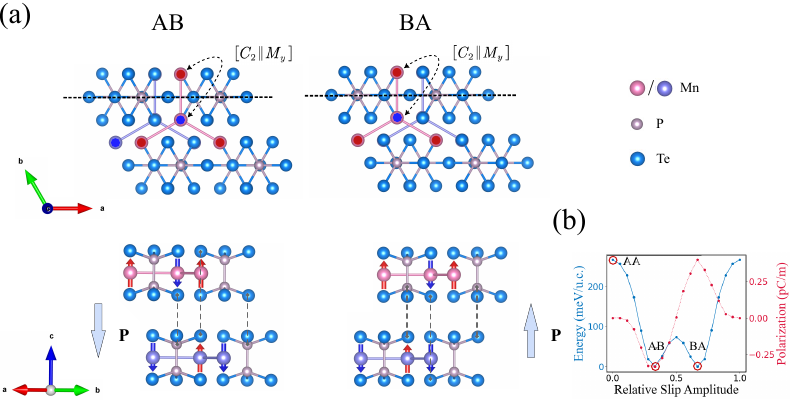}
\caption{ Stacking-dependent ferroelectricity in bilayer MnPTe$_3$. (a) Atomic structures of AB (left) and BA (right) stacking configuration. (b) Energy profiles (blue line) and the corresponding variation of out-of-plane electric polarization (red line) as functions of interlayer sliding along the [100] direction. }\label{stacking}
\end{figure*}

The broken in-plane mirror symmetry in polar stackings simultaneously lifts both valley degeneracy and \(\mathcal{PT}\)-protected spin degeneracy, creating an ideal platform for electrically controllable valley-spin-layer locking. As shown in Fig.~3, the SOC band structures for both AB- and BA-stacked MnPTe$_3$ bilayers reveal striking effects: AB stacking exhibits spontaneous valley polarization in both conduction and valence band, where the \(K'\)-valley CBM is 22.3~meV lower than the \(K\)-valley CBM, while the \(K\)-valley VBM sits 15.4~meV above its \(K'\)-valley counterpart (Fig. 3(a)). These substantial splittings are larger than those achieved through magnetic proximity effects and comparable to the values reported for sliding FE systems like bilayer T-FeCl$_2$ \cite{MaYandong-2022-FE_valley_bi_FeCl2-npj_CM} and YI$_2$ \cite{ZhangXianmin-2023-FE_valley_bi_YI2-NanoLetter}. Remarkably, polarization reversal (AB $\leftrightarrow$ BA) via interlayer sliding inverts valley polarizations (\(K' \leftrightarrow K\)) while maintaining splitting magnitudes (Fig. 3(b)), exactly as predicted by our \(k \cdot p\) model. Notably, while the global VBM resides at \(\Gamma\), the valley-polarized CBM (\(K\)/\(K'\)) states show distinct layer selectivity: the \(K'\) valley in AB stacking is dominated by lower-layer states, while the \(K\) valley in BA stacking is primarily contributed by upper-layer states. The strong coupling between layer-resolved valley polarization and FE states makes MnPTe$_3$ as a unique platform for practical valleytronic applications where spatial control of valley carriers is essential. 


\begin{figure*}
\includegraphics[scale=0.5]{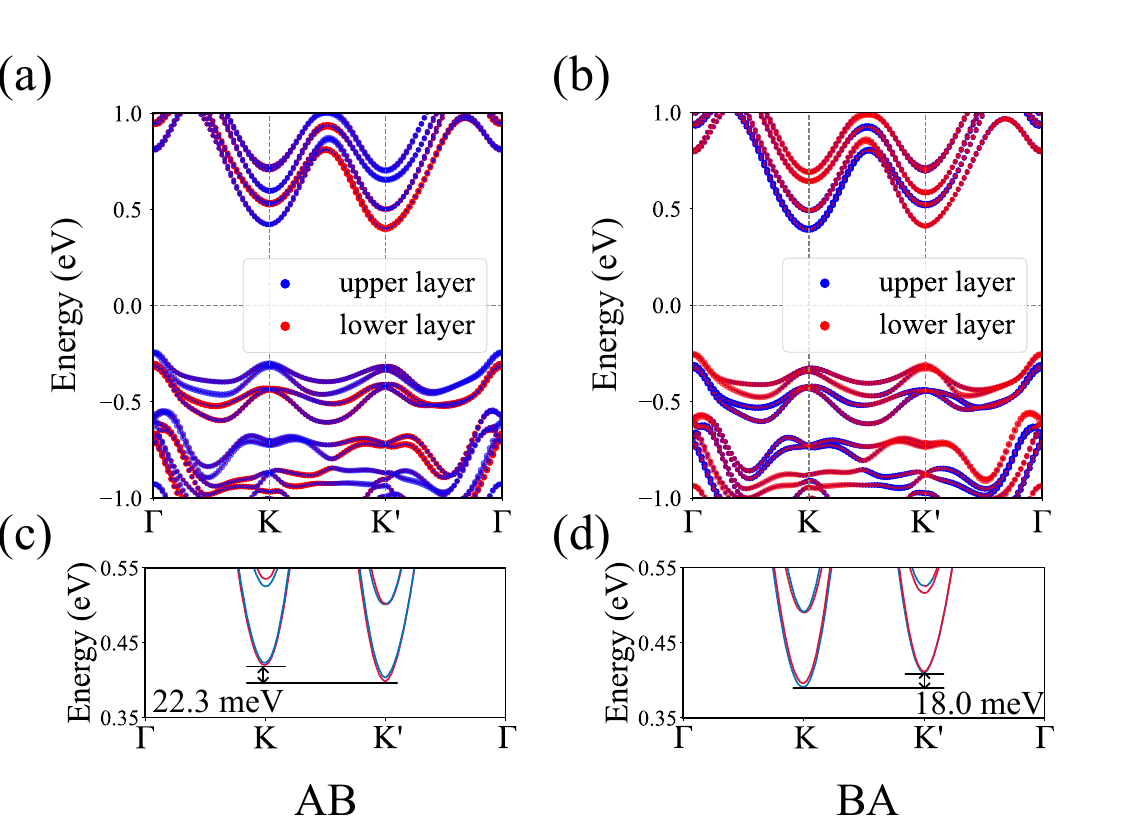}
\caption{Layer-projected electronic structure of bilayer MnPTe$_3$. (a)-(b) Band structures with SOC for AB and BA stacking configurations, respectively. The dot colors indicate layer contributions: blue (upper layer) and red (lower layer), (c)-(d) Zoomed-in views of the conduction band minima near the \(K\)/\(K'\) points from (a) and (b), showing out-of-plane spin polarization (red: spin-up, blue: spin-down). The spin-valley-layer locking manifests as opposite spin polarizations between AB and BA stackings, demonstrating electrically switchable spin textures through polarization reversal.
}\label{band_soc}
\end{figure*}

Consistent with our \(k \cdot p\) model predictions, Fig. 3(c,d) shows significant spin splitting near the \(K\) and \(K'\) valleys, in contrast to the spin-degenerate monolayer case. This emergent spin polarization arises from the broken \(\mathcal{PT}\) symmetry in polar-stacked bilayers and exhibits a remarkable layer-valley-spin locking effect. Specifically, the lower-layer-dominated \(K'\) valley CBM in AB stacking shows dominant spin-up character, while the upper-layer-dominated \(K\) valley CBM in BA stacking exhibits spin-down polarization. This demonstrates an unprecedented triple coupling between valley, spin, and layer degrees of freedom. Most remarkably, a simple interlayer sliding operation simultaneously inverts the FE polarization, valley polarization, and spin polarization while preserving their energy splittings, enabling fully reversible electrical control of layer-locked spin-polarized transport. Such deterministic spin-valley-layer locking, with its non-volatile switching characteristics, offers exceptional potential for developing programmable spintronic devices with multifunctional control capabilities.


Our parameterized \(k \cdot p\) model, carefully fitted to DFT results near the Fermi level at the \(K\) and \(K'\) points  (Table S1), well reproduces the spin-split band structure (Fig. 4). To elucidate the system's topological properties, we calculate the Berry curvature using the fitted model, which is defined as:
\[
\Omega_k^n = -2\sum_n \sum_{n \neq n'} \frac{ \, \text{Im} \langle \Psi_{nk} | v_x | \Psi_{n'k} \rangle \langle \Psi_{n'k} | v_y | \Psi_{nk} \rangle}{(E_n - E_{n'})^2}
\]
where \(|\Psi_{nk}\rangle\) represents the Bloch wave function with eigenvalue \(E_n\), and \(v_{x(y)}\) denotes the velocity operator. The calculated Berry curvatures for AB and BA stackings are given in Figs. 4(c) and 4(d), respectively. It can be seen that AB stacking exhibits same-sign Berry curvature at \(K\) and \(K'\) points, while BA stacking shows opposite signs. This distinctive pattern arises from valley nesting, where \(K\) and \(K'\) valleys feature bands with identical spin polarization but opposite layer character. These findings, further validated by independent VASPBERRY calculations (Fig. S3 in Supporting Information), demonstrate a unique Berry curvature signature that  distinguishes our system from previous reports\cite{Feng-2025-FE_valley_alter_PT-PRB}.

Given the fact that the interlayer sliding-induced FE polarization reversal inverts both the CBM spin polarization and the Berry curvature sign, a layer- and spin-locked anomalous valley Hall effect (AVHE) can be expected under moderate n-type doping. When the Fermi level lies within the energy window between \(K\) and \(K'\) valleys, the system exhibits striking polarization-dependent behavior. As shown in Fig. 4(c), for downward FE polarization, spin-up electrons at the \(K'\) valley acquire an anomalous velocity \(v_a \sim E \times \Omega_k^n\), which is proportional to the negative Berry curvature, causing them to accumulate on the right side of the lower layer. When the FE polarization is reversed (Fig. 4(d)), spin-down electrons at the \(K\) valley now experience opposite anomalous velocity, resulting left-side accumulation in the upper layer. This behavior is further corroborated by the calculated anomalous Hall conductivity , which is given as
\[
\sigma _{xy}=-\frac{e^2}{\hbar}\sum_{k,n}{\Omega _{k}^{n}f\left( \varepsilon _{k}^{n}-\varepsilon _F \right)},
\]
where \(f\left( \varepsilon _{k}^{n}-\varepsilon _F \right)\) is the Fermi–Dirac distribution function. As shown in Fig.~4(e), the conductivity undergoes a pronounced sign reversal near the CBM when switching between the two stacking configurations, directly mirroring the Berry curvature inversion. These results suggest nonvolatile control of valley-spin-layer coupled transport through sliding ferroelectricity, opening new possibilities for novel quantum devices that exploit the coherent manipulation of multiple electronic  degrees of freedom.

\begin{figure}
\includegraphics[scale=0.65]{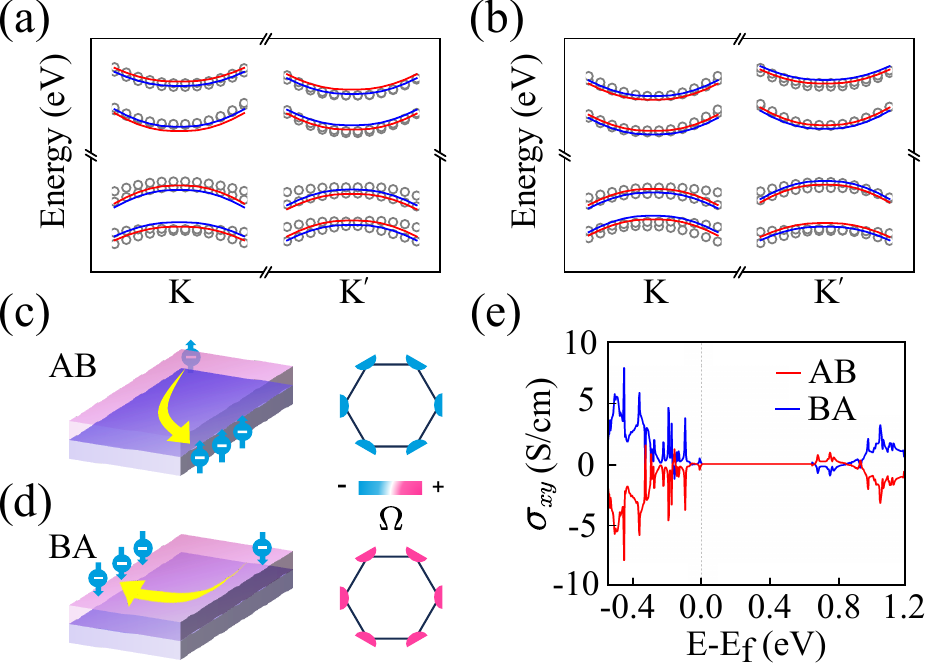}
\caption{ Electronic structure and topological transport properties of bilayer MnPTe$_3$. (a)-(b) Band structures near the Fermi level for AB- and BA-stacked bilayer MnPTe$_3$ with SOC, comparing DFT results (gray circles) with \(k\cdot p\) model fits (red/blue lines for opposite spin polarizations. (c)-(d) Left: Schematic of the spin--layer--polarized anomalous valley Hall effect at the CBM, showing opposite carrier accumulation for each stacking. Right: Calculated Berry curvature distributions around $K$ and $K'$ valleys, demonstrating stacking-dependent signatures: AB shows identical curvature at \(K/K'\) valleys (red lobes), while BA exhibits sign-inverted patterns (blue lobes). 
(e) Anomalous Hall conductivity $\sigma _{xy}$ computed via WannierTools, demonstrating sign reversal between stacking configurations due to Berry curvature inversion.
}\label{BC}
\end{figure}

\begin{figure*}
\includegraphics[scale=0.6]{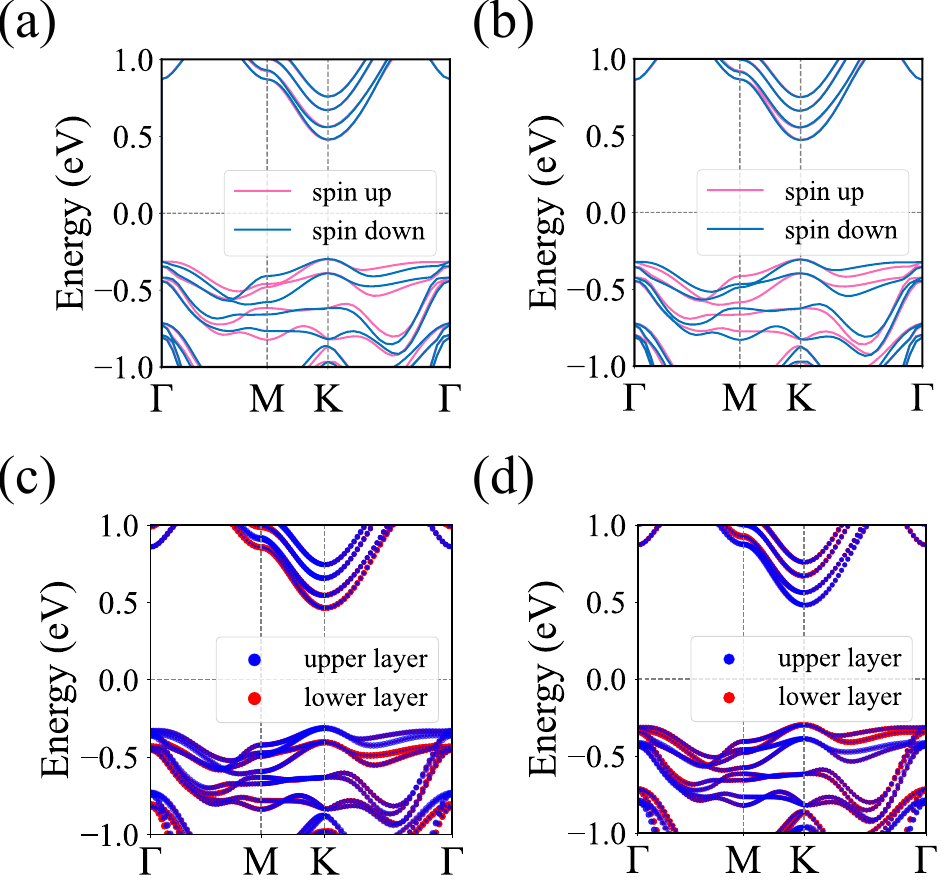}
\caption{ (a)-(b) Band structures of AB- and BA-stacked bilayer MnPTe$_3$ without SOC. (c)-(d) Projected band structures of AB- and BA-stacked bilayer MnPTe$_3$ without SOC, where red and blue dots indicate the contributions from the lower and upper layers, respectively.}\label{band_soc}
\end{figure*}

Besides enabling nonvolatile control of AVHE, bilayer MnPTe$_3$ also exhibits robust altermagnetism, which is absent in its  monolayer counterpart. While the monolayer maintains \(\mathcal{PT}\) symmetry, preventing 
spin splitting throughout the Brillouin zone, the polar bilayer breaks \(\mathcal{PT}\) symmetry and only one \(\mathcal{M T}\) symmetry survives that connects spin-up and spin-down sublattices (see Fig. 2(a)). This symmetry reduction permits momentum-dependent spin-splitting while maintaining zero net magnetization, a hallmark of altermagnetism. Fig. 5(a) shows the spin-polarized band structure without SOC for the AB stacking bilayer MnPTe$_3$. Large nonrelativistic spin splittings can be seen along the paths that violate the mirror symmetry, particularly in the valence bands. It is interesting to note that these splittings show layer-resolved inversion: the upper layer's spin splitting is reverted relative to the lower layer's due to the stacking operation's inherent time-reversal \(\mathcal{T}\) connection between layers (Fig. S4). Reversing the electric polarization (AB $\rightarrow$ BA) switches the dominant layer in the valence bands and inverts the spin splitting along paths like \(\Gamma-K\) (Fig. 5(b)), without altering the Mn ions' magnetic moments in the system\cite{Sun-2025-alter,Sun-2025-stacking_alter}. This realizes the altermagnetic magnetoelectric effect, fundamentally different from conventional multiferroic coupling. This polarization-driven switching of spin splittings occurs without magnetic moment reorientation, enabling ultrafast manipulation of spin-polarized states, a critical advantage for high-speed spintronic devices requiring low-energy switching. 


In summary, we develop a four-band spin-full low-energy effective \(k \cdot p\) model to elucidate the coupling mechanisms between ferroelectricity, valley polarization, and altermagnetism in intralayer AFM systems with honeycomb lattices. This general design principle is subsequently validated in real materials via first-principles calculations. Using bilayer MnPTe$_3$ as a representative example, we demonstrate that interlayer sliding can control layer-resolved valley polarization and altermagnetic spin splitting through FE polarization reversal. Our work highlights the potential for achieving electrical control over valley, spin, and layer degrees of freedom in two-dimensional antiferromagnets, opening new avenues for multifunctional device applications that integrate valleytronics, spintronics, and layertronics. Importantly, MnPTe$_3$ is not the sole candidate for realizing this mechanism. Similar effects are expected in other honeycomb-structured intralayer antiferromagnets, such as strained CrCl$_3$(see Fig. S6-S8), thereby providing a broader materials platform for the exploration and engineering of electrically tunable quantum functionalities.

\section{Computational details}
The first-principles calculations are performed within the framework of density functional theory (DFT) using the projected augmented wave (PAW) method, as implemented in the Vienna Ab initio Simulation Package (VASP)\cite{Blochl-1994-PAW,kresse-1994-GGA,kresse-1996-efficient-iterative-schemes} . The exchange-correlation functional is treated using the generalized gradient approximation (GGA) of the Perdew-Burke-Ernzerhof (PBE) functional\cite{perdew-1996-PBE,kresse-1999-PAW}. Kohn-Sham orbitals are expanded in a plane-wave basis set with an energy cutoff of 500 eV. To account for electron correlation in the localized $d$ orbitals, an effective on-site Hubbard $U$ of 5.0\,eV is applied for Mn atoms~\cite{MacDonald-2016-MPX3-PRB,Fangping-2023-MnPTe3-PRB}, and 3\,eV for Cr atoms\cite{huang-2018-CrX3prediction-PRL,xue-2019-CrCl3-PRB}. All structures are fully relaxed until the atomic forces are below $10^{-2}$ eV/{\AA} and the energy difference between consecutive self-consistent steps is less than $10^{-5}$ eV. To sample the Brillouin zone (BZ), we use a 9 × 9 × 1  $\Gamma$-centered Monkhorst-Pack $k$-point grid. A vacuum spacing of 20 {\AA} is included along the non-periodic direction to prevent interactions between adjacent images. The vdW interactions between layers are described using the DFT-D3 method\cite{Grimme-2011-DFT_D3-JCC}. Convergence tests have been performed with respect to the energy cutoff and $k$-point sampling. Berry curvatures are computed using VASPBERRY~\cite{Kim-2022-VASPBERRY}, and the anomalous Hall conductivity is evaluated using WannierTools~\cite{WannierTools}, which is a post-processing tool based on maximally localized Wannier functions (MLWFs) generated by the Wannier90 package~\cite{Wannier90}.

\begin{acknowledgement}

This work was supported by the National Key R\&D Program of China (nos. 2022YFA1402903 and 2024YFA1409101), the National 
Natural Science Foundation of China under Grant Nos. 12374059 and 12004160, and the Shenzhen Science and Technology Program (Grant No. JCYJ20240813095301003). Z.G. also acknowledges the financial support from the Shenzhen Science and Technology 
Program (Grant KQTD20190929173815000) and the Guangdong Innovative and Entrepreneurial Research Team Program (Grant 2019ZT08C044). Computational time was supported by the Center for Computational Science and Engineering of Southern University 
of Science and Technology and the Major Science and Technology Infrastructure Project of Material Genome Big-science 
Facilities Platform supported by Municipal Development and Reform Commission of Shenzhen.

\end{acknowledgement}

\begin{suppinfo}
Detailed theoretical analysis of spin-valley-layer coupling and altermagnetism based on \(k \cdot p\) models; model parameters fitted to DFT results; additional DFT data including band structures, spin splittings, Berry curvature, and structural information for bilayer MnPTe\(_3\) and strained CrCl\(_3\) (PDF).

\begin{itemize}
  \item \texttt{Supp.pdf}
\end{itemize}
\end{suppinfo}

\bibliography{achemso-demo}

\providecommand{\latin}[1]{#1}
\makeatletter
\providecommand{\doi}
  {\begingroup\let\do\@makeother\dospecials
  \catcode`\{=1 \catcode`\}=2 \doi@aux}
\providecommand{\doi@aux}[1]{\endgroup\texttt{#1}}
\makeatother
\providecommand*\mcitethebibliography{\thebibliography}
\csname @ifundefined\endcsname{endmcitethebibliography}  {\let\endmcitethebibliography\endthebibliography}{}
\begin{mcitethebibliography}{61}
\providecommand*\natexlab[1]{#1}
\providecommand*\mciteSetBstSublistMode[1]{}
\providecommand*\mciteSetBstMaxWidthForm[2]{}
\providecommand*\mciteBstWouldAddEndPuncttrue
  {\def\EndOfBibitem{\unskip.}}
\providecommand*\mciteBstWouldAddEndPunctfalse
  {\let\EndOfBibitem\relax}
\providecommand*\mciteSetBstMidEndSepPunct[3]{}
\providecommand*\mciteSetBstSublistLabelBeginEnd[3]{}
\providecommand*\EndOfBibitem{}
\mciteSetBstSublistMode{f}
\mciteSetBstMaxWidthForm{subitem}{(\alph{mcitesubitemcount})}
\mciteSetBstSublistLabelBeginEnd
  {\mcitemaxwidthsubitemform\space}
  {\relax}
  {\relax}

\bibitem[Xiao \latin{et~al.}(2007)Xiao, Yao, and Niu]{XiaoDi-2007-valley-PRL}
Xiao,~D.; Yao,~W.; Niu,~Q. Valley-contrasting physics in graphene: magnetic moment and topological transport. \emph{Phys. Rev. Lett.} \textbf{2007}, \emph{99}, 236809\relax
\mciteBstWouldAddEndPuncttrue
\mciteSetBstMidEndSepPunct{\mcitedefaultmidpunct}
{\mcitedefaultendpunct}{\mcitedefaultseppunct}\relax
\EndOfBibitem
\bibitem[Yao \latin{et~al.}(2008)Yao, Xiao, and Niu]{YaoWang-2008-Valley_optelectronics-PRB}
Yao,~W.; Xiao,~D.; Niu,~Q. Valley-dependent optoelectronics from inversion symmetry breaking. \emph{Phys. Rev. B} \textbf{2008}, \emph{77}, 235406\relax
\mciteBstWouldAddEndPuncttrue
\mciteSetBstMidEndSepPunct{\mcitedefaultmidpunct}
{\mcitedefaultendpunct}{\mcitedefaultseppunct}\relax
\EndOfBibitem
\bibitem[Xiao \latin{et~al.}(2012)Xiao, Liu, Feng, Xu, and Yao]{XiaoDi-2012-valley_MoS2-PRL}
Xiao,~D.; Liu,~G.-B.; Feng,~W.; Xu,~X.; Yao,~W. Coupled Spin and Valley Physics in Monolayers of MoS$_{2}$ and Other Group-VI Dichalcogenides. \emph{Phys. Rev. Lett.} \textbf{2012}, \emph{108}, 196802\relax
\mciteBstWouldAddEndPuncttrue
\mciteSetBstMidEndSepPunct{\mcitedefaultmidpunct}
{\mcitedefaultendpunct}{\mcitedefaultseppunct}\relax
\EndOfBibitem
\bibitem[Schaibley \latin{et~al.}(2016)Schaibley, Yu, Clark, Rivera, Ross, Seyler, Yao, and Xu]{Schaibley-2016-valleytronics_Review-NRM}
Schaibley,~J.~R.; Yu,~H.; Clark,~G.; Rivera,~P.; Ross,~J.~S.; Seyler,~K.~L.; Yao,~W.; Xu,~X. Valleytronics in 2D materials. \emph{Nat. Rev. Mater.} \textbf{2016}, \emph{1}, 1--15\relax
\mciteBstWouldAddEndPuncttrue
\mciteSetBstMidEndSepPunct{\mcitedefaultmidpunct}
{\mcitedefaultendpunct}{\mcitedefaultseppunct}\relax
\EndOfBibitem
\bibitem[Vitale \latin{et~al.}(2018)Vitale, Nezich, Varghese, Kim, Gedik, Jarillo-Herrero, Xiao, and Rothschild]{Vitale-2018-valleytronics_Review-Small}
Vitale,~S.~A.; Nezich,~D.; Varghese,~J.~O.; Kim,~P.; Gedik,~N.; Jarillo-Herrero,~P.; Xiao,~D.; Rothschild,~M. Valleytronics: opportunities, challenges, and paths forward. \emph{Small} \textbf{2018}, \emph{14}, 1801483\relax
\mciteBstWouldAddEndPuncttrue
\mciteSetBstMidEndSepPunct{\mcitedefaultmidpunct}
{\mcitedefaultendpunct}{\mcitedefaultseppunct}\relax
\EndOfBibitem
\bibitem[Pacchioni(2020)]{Pacchioni-2020-valleytronics_comment-NRM}
Pacchioni,~G. Valleytronics with a twist. \emph{Nat. Rev. Mater.} \textbf{2020}, \emph{5}, 480--480\relax
\mciteBstWouldAddEndPuncttrue
\mciteSetBstMidEndSepPunct{\mcitedefaultmidpunct}
{\mcitedefaultendpunct}{\mcitedefaultseppunct}\relax
\EndOfBibitem
\bibitem[\ifmmode~\check{S}\else \v{S}\fi{}mejkal \latin{et~al.}(2022)\ifmmode~\check{S}\else \v{S}\fi{}mejkal, Sinova, and Jungwirth]{Smejkal-2022-alter1-PRX}
\ifmmode~\check{S}\else \v{S}\fi{}mejkal,~L.; Sinova,~J.; Jungwirth,~T. Emerging Research Landscape of Altermagnetism. \emph{Phys. Rev. X} \textbf{2022}, \emph{12}, 040501\relax
\mciteBstWouldAddEndPuncttrue
\mciteSetBstMidEndSepPunct{\mcitedefaultmidpunct}
{\mcitedefaultendpunct}{\mcitedefaultseppunct}\relax
\EndOfBibitem
\bibitem[\ifmmode~\check{S}\else \v{S}\fi{}mejkal \latin{et~al.}(2022)\ifmmode~\check{S}\else \v{S}\fi{}mejkal, Sinova, and Jungwirth]{Smejkal-2022-alter2-PRX}
\ifmmode~\check{S}\else \v{S}\fi{}mejkal,~L.; Sinova,~J.; Jungwirth,~T. Beyond Conventional Ferromagnetism and Antiferromagnetism: A Phase with Nonrelativistic Spin and Crystal Rotation Symmetry. \emph{Phys. Rev. X} \textbf{2022}, \emph{12}, 031042\relax
\mciteBstWouldAddEndPuncttrue
\mciteSetBstMidEndSepPunct{\mcitedefaultmidpunct}
{\mcitedefaultendpunct}{\mcitedefaultseppunct}\relax
\EndOfBibitem
\bibitem[Zhu \latin{et~al.}(2024)Zhu, Chen, Liu, Liu, Liu, Zha, Qu, Hong, Li, Jiang, \latin{et~al.} others]{zhu-2024-alter_observation_nature}
Zhu,~Y.-P.; Chen,~X.; Liu,~X.-R.; Liu,~Y.; Liu,~P.; Zha,~H.; Qu,~G.; Hong,~C.; Li,~J.; Jiang,~Z.; others Observation of plaid-like spin splitting in a noncoplanar antiferromagnet. \emph{Nature} \textbf{2024}, \emph{626}, 523--528\relax
\mciteBstWouldAddEndPuncttrue
\mciteSetBstMidEndSepPunct{\mcitedefaultmidpunct}
{\mcitedefaultendpunct}{\mcitedefaultseppunct}\relax
\EndOfBibitem
\bibitem[Bai \latin{et~al.}(2024)Bai, Feng, Liu, {\v{S}}mejkal, Mokrousov, and Yao]{bai-2024-alter-AFM}
Bai,~L.; Feng,~W.; Liu,~S.; {\v{S}}mejkal,~L.; Mokrousov,~Y.; Yao,~Y. Altermagnetism: Exploring new frontiers in magnetism and spintronics. \emph{Adv. Funct. Mater.} \textbf{2024}, \emph{34}, 2409327\relax
\mciteBstWouldAddEndPuncttrue
\mciteSetBstMidEndSepPunct{\mcitedefaultmidpunct}
{\mcitedefaultendpunct}{\mcitedefaultseppunct}\relax
\EndOfBibitem
\bibitem[Mak \latin{et~al.}(2014)Mak, McGill, Park, and McEuen]{mak-2014-MoS2-Science}
Mak,~K.~F.; McGill,~K.~L.; Park,~J.; McEuen,~P.~L. The valley Hall effect in MoS$_{2}$ transistors. \emph{Science} \textbf{2014}, \emph{344}, 1489--1492\relax
\mciteBstWouldAddEndPuncttrue
\mciteSetBstMidEndSepPunct{\mcitedefaultmidpunct}
{\mcitedefaultendpunct}{\mcitedefaultseppunct}\relax
\EndOfBibitem
\bibitem[Lee \latin{et~al.}(2016)Lee, Mak, and Shan]{lee-2016-bilayer_MoS2-Nature_nano}
Lee,~J.; Mak,~K.~F.; Shan,~J. Electrical control of the valley Hall effect in bilayer MoS$_{2}$ transistors. \emph{Nat. Nanotechnol.} \textbf{2016}, \emph{11}, 421--425\relax
\mciteBstWouldAddEndPuncttrue
\mciteSetBstMidEndSepPunct{\mcitedefaultmidpunct}
{\mcitedefaultendpunct}{\mcitedefaultseppunct}\relax
\EndOfBibitem
\bibitem[Wu \latin{et~al.}(2024)Wu, Deng, Yin, Tong, Tian, and Zhang]{Wu-2024-valley_alter-nanoletter}
Wu,~Y.; Deng,~L.; Yin,~X.; Tong,~J.; Tian,~F.; Zhang,~X. Valley-Related Multipiezo Effect and Noncollinear Spin Current in an Altermagnet Fe$_2$Se$_2$O Monolayer. \emph{Nano Lett.} \textbf{2024}, \emph{24}, 10534--10539, PMID: 39145607\relax
\mciteBstWouldAddEndPuncttrue
\mciteSetBstMidEndSepPunct{\mcitedefaultmidpunct}
{\mcitedefaultendpunct}{\mcitedefaultseppunct}\relax
\EndOfBibitem
\bibitem[Guo \latin{et~al.}(2024)Guo, Guo, and Wang]{guo-2024-valley_alter-2024}
Guo,~S.-D.; Guo,~X.-S.; Wang,~G. Valley polarization in two-dimensional tetragonal altermagnetism. \emph{Phys. Rev. B} \textbf{2024}, \emph{110}, 184408\relax
\mciteBstWouldAddEndPuncttrue
\mciteSetBstMidEndSepPunct{\mcitedefaultmidpunct}
{\mcitedefaultendpunct}{\mcitedefaultseppunct}\relax
\EndOfBibitem
\bibitem[Feng \latin{et~al.}(2025)Feng, Zhou, Chen, Xu, Yang, and Li]{Feng-2025-FE_valley_alter_PT-PRB}
Feng,~J.; Zhou,~X.; Chen,~J.; Xu,~M.; Yang,~X.; Li,~Y. Ferroelectric antiferromagnetic lifting of spin-valley degeneracy. \emph{Phys. Rev. B} \textbf{2025}, \emph{111}, 214446\relax
\mciteBstWouldAddEndPuncttrue
\mciteSetBstMidEndSepPunct{\mcitedefaultmidpunct}
{\mcitedefaultendpunct}{\mcitedefaultseppunct}\relax
\EndOfBibitem
\bibitem[Zheng \latin{et~al.}(2022)Zheng, Zhao, Tan, Guan, Zhong, Yue, Xiang, and Duan]{DuanChungang-2022-FE_valley_coupling_review-APL}
Zheng,~J.-D.; Zhao,~Y.-F.; Tan,~Y.-F.; Guan,~Z.; Zhong,~N.; Yue,~F.-Y.; Xiang,~P.-H.; Duan,~C.-G. Coupling of ferroelectric and valley properties in 2D materials. \emph{J. Appl. Phys.} \textbf{2022}, \emph{132}\relax
\mciteBstWouldAddEndPuncttrue
\mciteSetBstMidEndSepPunct{\mcitedefaultmidpunct}
{\mcitedefaultendpunct}{\mcitedefaultseppunct}\relax
\EndOfBibitem
\bibitem[Pei \latin{et~al.}(2019)Pei, Zhou, Mi, and Cheng]{PeiQi-2019-FE_valley_AgBiP2Se6/CrI3-ACS_AMI}
Pei,~Q.; Zhou,~B.; Mi,~W.; Cheng,~Y. Triferroic Material and Electrical Control of Valley Degree of Freedom. \emph{ACS Appl. Mater. Interfaces} \textbf{2019}, \emph{11}, 12675--12682\relax
\mciteBstWouldAddEndPuncttrue
\mciteSetBstMidEndSepPunct{\mcitedefaultmidpunct}
{\mcitedefaultendpunct}{\mcitedefaultseppunct}\relax
\EndOfBibitem
\bibitem[Hu \latin{et~al.}(2020)Hu, Tong, Shen, and Duan]{DuanChungang-2020-FE_valley_CuInP2S6/MnPSe3-JMCC}
Hu,~H.; Tong,~W.-Y.; Shen,~Y.-H.; Duan,~C.-G. Electrical control of the valley degree of freedom in 2D ferroelectric/antiferromagnetic heterostructures. \emph{J. Mater. Chem. C} \textbf{2020}, \emph{8}, 8098--8106\relax
\mciteBstWouldAddEndPuncttrue
\mciteSetBstMidEndSepPunct{\mcitedefaultmidpunct}
{\mcitedefaultendpunct}{\mcitedefaultseppunct}\relax
\EndOfBibitem
\bibitem[Yin \latin{et~al.}(2018)Yin, Wang, and Mi]{MiWenbo-2018-AFE_valley_BiXO3/BiIrO3-ACS_AMI}
Yin,~L.; Wang,~X.; Mi,~W. Tunable Valley and Spin Polarizations in BiXO$_3$/BiIrO$_3$ (X = Fe, Mn) Ferroelectric Superlattices. \emph{ACS Appl. Mater. Interfaces} \textbf{2018}, \emph{10}, 3822--3829, PMID: 29322771\relax
\mciteBstWouldAddEndPuncttrue
\mciteSetBstMidEndSepPunct{\mcitedefaultmidpunct}
{\mcitedefaultendpunct}{\mcitedefaultseppunct}\relax
\EndOfBibitem
\bibitem[Ma \latin{et~al.}(2020)Ma, Shao, Fan, Liu, Feng, Sun, and Zhao]{MaXikui-2020-FE_valley_VTe2/Ga2S3-JMCC}
Ma,~X.; Shao,~X.; Fan,~Y.; Liu,~J.; Feng,~X.; Sun,~L.; Zhao,~M. Tunable valley splitting and anomalous valley Hall effect in VTe $_2$/Ga$_2$S$_3$ heterostructures. \emph{J. Mater. Chem. C} \textbf{2020}, \emph{8}, 14895--14901\relax
\mciteBstWouldAddEndPuncttrue
\mciteSetBstMidEndSepPunct{\mcitedefaultmidpunct}
{\mcitedefaultendpunct}{\mcitedefaultseppunct}\relax
\EndOfBibitem
\bibitem[Zhai \latin{et~al.}(2021)Zhai, Cheng, Yao, Yin, Shen, Xia, and He]{zhai2021using}
Zhai,~B.; Cheng,~R.; Yao,~W.; Yin,~L.; Shen,~C.; Xia,~C.; He,~J. Using ferroelectric polarization to regulate and preserve the valley polarization in a HfN$_2$/CrI$_3$/In$_2$Se$_3$ heterotrilayer. \emph{Phys. Rev. B} \textbf{2021}, \emph{103}, 214114\relax
\mciteBstWouldAddEndPuncttrue
\mciteSetBstMidEndSepPunct{\mcitedefaultmidpunct}
{\mcitedefaultendpunct}{\mcitedefaultseppunct}\relax
\EndOfBibitem
\bibitem[Lai \latin{et~al.}(2019)Lai, Song, Wan, Xue, Wang, Ye, Dai, Zhang, Yang, Du, \latin{et~al.} others]{lai2019two}
Lai,~Y.; Song,~Z.; Wan,~Y.; Xue,~M.; Wang,~C.; Ye,~Y.; Dai,~L.; Zhang,~Z.; Yang,~W.; Du,~H.; others Two-dimensional ferromagnetism and driven ferroelectricity in van der Waals CuCrP 2 S 6. \emph{Nanoscale} \textbf{2019}, \emph{11}, 5163--5170\relax
\mciteBstWouldAddEndPuncttrue
\mciteSetBstMidEndSepPunct{\mcitedefaultmidpunct}
{\mcitedefaultendpunct}{\mcitedefaultseppunct}\relax
\EndOfBibitem
\bibitem[Shen \latin{et~al.}(2017)Shen, Tong, Gong, and Duan]{DuanChungang-2017-FE_valley_GeSe-2DMaterials}
Shen,~X.-W.; Tong,~W.-Y.; Gong,~S.-J.; Duan,~C.-G. Electrically tunable polarizer based on 2D orthorhombic ferrovalley materials. \emph{2D Mater.} \textbf{2017}, \emph{5}, 011001\relax
\mciteBstWouldAddEndPuncttrue
\mciteSetBstMidEndSepPunct{\mcitedefaultmidpunct}
{\mcitedefaultendpunct}{\mcitedefaultseppunct}\relax
\EndOfBibitem
\bibitem[Xiao \latin{et~al.}(2018)Xiao, Wang, Yang, Lu, Feng, and Zhang]{XiaoChengcheng-2018-FE_valley_Bi-AdvancedFM}
Xiao,~C.; Wang,~F.; Yang,~S.~A.; Lu,~Y.; Feng,~Y.; Zhang,~S. Elemental ferroelectricity and antiferroelectricity in Group-V monolayer. \emph{Adv. Funct. Mater.} \textbf{2018}, \emph{28}, 1707383\relax
\mciteBstWouldAddEndPuncttrue
\mciteSetBstMidEndSepPunct{\mcitedefaultmidpunct}
{\mcitedefaultendpunct}{\mcitedefaultseppunct}\relax
\EndOfBibitem
\bibitem[Li \latin{et~al.}(2019)Li, Wu, and Zeng]{LiLei-2019-AdsorpFE_valley_XCl2(C2N)6-AdvancedFM}
Li,~L.; Wu,~M.; Zeng,~X.~C. Facile and versatile functionalization of two-dimensional carbon nitrides by design: magnetism/multiferroicity, valleytronics, and photovoltaics. \emph{Adv. Funct. Mater.} \textbf{2019}, \emph{29}, 1905752\relax
\mciteBstWouldAddEndPuncttrue
\mciteSetBstMidEndSepPunct{\mcitedefaultmidpunct}
{\mcitedefaultendpunct}{\mcitedefaultseppunct}\relax
\EndOfBibitem
\bibitem[Liu \latin{et~al.}(2020)Liu, Pyatakov, and Ren]{LiuXingen-2020-FE_valley_bi_VS2-PRL}
Liu,~X.; Pyatakov,~A.~P.; Ren,~W. Magnetoelectric coupling in multiferroic bilayer VS$_2$. \emph{Phys. Rev. Lett.} \textbf{2020}, \emph{125}, 247601\relax
\mciteBstWouldAddEndPuncttrue
\mciteSetBstMidEndSepPunct{\mcitedefaultmidpunct}
{\mcitedefaultendpunct}{\mcitedefaultseppunct}\relax
\EndOfBibitem
\bibitem[Zhang \latin{et~al.}(2022)Zhang, Xu, Huang, Dai, and Ma]{MaYandong-2022-FE_valley_bi_FeCl2-npj_CM}
Zhang,~T.; Xu,~X.; Huang,~B.; Dai,~Y.; Ma,~Y. 2D spontaneous valley polarization from inversion symmetric single-layer lattices. \emph{npj Comput. Mater.} \textbf{2022}, \emph{8}, 64\relax
\mciteBstWouldAddEndPuncttrue
\mciteSetBstMidEndSepPunct{\mcitedefaultmidpunct}
{\mcitedefaultendpunct}{\mcitedefaultseppunct}\relax
\EndOfBibitem
\bibitem[Wu \latin{et~al.}(2023)Wu, Tong, Deng, Luo, Tian, Qin, and Zhang]{ZhangXianmin-2023-FE_valley_bi_YI2-NanoLetter}
Wu,~Y.; Tong,~J.; Deng,~L.; Luo,~F.; Tian,~F.; Qin,~G.; Zhang,~X. Coexisting ferroelectric and ferrovalley polarizations in bilayer stacked magnetic semiconductors. \emph{Nano Lett.} \textbf{2023}, \emph{23}, 6226--6232\relax
\mciteBstWouldAddEndPuncttrue
\mciteSetBstMidEndSepPunct{\mcitedefaultmidpunct}
{\mcitedefaultendpunct}{\mcitedefaultseppunct}\relax
\EndOfBibitem
\bibitem[Gui and Huang(2025)Gui, and Huang]{Gui_2025}
Gui,~Z.; Huang,~L. Stacking ferroelectricity in two-dimensional van der Waals materials. \emph{J. Phys. Condens. Matter} \textbf{2025}, \emph{37}, 113005\relax
\mciteBstWouldAddEndPuncttrue
\mciteSetBstMidEndSepPunct{\mcitedefaultmidpunct}
{\mcitedefaultendpunct}{\mcitedefaultseppunct}\relax
\EndOfBibitem
\bibitem[{\v{S}}mejkal \latin{et~al.}(2020){\v{S}}mejkal, Gonz{\'a}lez-Hern{\'a}ndez, Jungwirth, and Sinova]{vsmejkal-2020-RuO2}
{\v{S}}mejkal,~L.; Gonz{\'a}lez-Hern{\'a}ndez,~R.; Jungwirth,~T.; Sinova,~J. Crystal time-reversal symmetry breaking and spontaneous Hall effect in collinear antiferromagnets. \emph{Sci. Adv.} \textbf{2020}, \emph{6}, eaaz8809\relax
\mciteBstWouldAddEndPuncttrue
\mciteSetBstMidEndSepPunct{\mcitedefaultmidpunct}
{\mcitedefaultendpunct}{\mcitedefaultseppunct}\relax
\EndOfBibitem
\bibitem[Bai \latin{et~al.}(2022)Bai, Han, Feng, Zhou, Su, Wang, Liao, Zhu, Chen, Pan, \latin{et~al.} others]{Bai-2022-RuO2}
Bai,~H.; Han,~L.; Feng,~X.; Zhou,~Y.; Su,~R.; Wang,~Q.; Liao,~L.; Zhu,~W.; Chen,~X.; Pan,~F.; others Observation of spin splitting torque in a collinear antiferromagnet RuO$_2$. \emph{Phys. Rev. Lett.} \textbf{2022}, \emph{128}, 197202\relax
\mciteBstWouldAddEndPuncttrue
\mciteSetBstMidEndSepPunct{\mcitedefaultmidpunct}
{\mcitedefaultendpunct}{\mcitedefaultseppunct}\relax
\EndOfBibitem
\bibitem[Lee \latin{et~al.}(2024)Lee, Lee, Jung, Jung, Kim, Lee, Seok, Kim, Park, {\v{S}}mejkal, \latin{et~al.} others]{Lee-2024-MnTe}
Lee,~S.; Lee,~S.; Jung,~S.; Jung,~J.; Kim,~D.; Lee,~Y.; Seok,~B.; Kim,~J.; Park,~B.~G.; {\v{S}}mejkal,~L.; others Broken Kramers degeneracy in altermagnetic MnTe. \emph{Phys. Rev. Lett.} \textbf{2024}, \emph{132}, 036702\relax
\mciteBstWouldAddEndPuncttrue
\mciteSetBstMidEndSepPunct{\mcitedefaultmidpunct}
{\mcitedefaultendpunct}{\mcitedefaultseppunct}\relax
\EndOfBibitem
\bibitem[Mazin(2023)]{Mazin-2023-MnTe}
Mazin,~I. Altermagnetism in MnTe: Origin, predicted manifestations, and routes to detwinning. \emph{Phys. Rev. B} \textbf{2023}, \emph{107}, L100418\relax
\mciteBstWouldAddEndPuncttrue
\mciteSetBstMidEndSepPunct{\mcitedefaultmidpunct}
{\mcitedefaultendpunct}{\mcitedefaultseppunct}\relax
\EndOfBibitem
\bibitem[Krempask{\`y} \latin{et~al.}(2024)Krempask{\`y}, {\v{S}}mejkal, D’souza, Hajlaoui, Springholz, Uhl{\'\i}{\v{r}}ov{\'a}, Alarab, Constantinou, Strocov, Usanov, \latin{et~al.} others]{krempasky-2024-MnTe}
Krempask{\`y},~J.; {\v{S}}mejkal,~L.; D’souza,~S.; Hajlaoui,~M.; Springholz,~G.; Uhl{\'\i}{\v{r}}ov{\'a},~K.; Alarab,~F.; Constantinou,~P.; Strocov,~V.; Usanov,~D.; others Altermagnetic lifting of Kramers spin degeneracy. \emph{Nature} \textbf{2024}, \emph{626}, 517--522\relax
\mciteBstWouldAddEndPuncttrue
\mciteSetBstMidEndSepPunct{\mcitedefaultmidpunct}
{\mcitedefaultendpunct}{\mcitedefaultseppunct}\relax
\EndOfBibitem
\bibitem[{\v{S}}mejkal \latin{et~al.}(2022){\v{S}}mejkal, Sinova, and Jungwirth]{vsmejkal-2022-CrSb}
{\v{S}}mejkal,~L.; Sinova,~J.; Jungwirth,~T. Beyond conventional ferromagnetism and antiferromagnetism: A phase with nonrelativistic spin and crystal rotation symmetry. \emph{Phys. Rev. X} \textbf{2022}, \emph{12}, 031042\relax
\mciteBstWouldAddEndPuncttrue
\mciteSetBstMidEndSepPunct{\mcitedefaultmidpunct}
{\mcitedefaultendpunct}{\mcitedefaultseppunct}\relax
\EndOfBibitem
\bibitem[Reimers \latin{et~al.}(2024)Reimers, Odenbreit, {\v{S}}mejkal, Strocov, Constantinou, Hellenes, Jaeschke~Ubiergo, Campos, Bharadwaj, Chakraborty, \latin{et~al.} others]{Reimers-2024-CrSb}
Reimers,~S.; Odenbreit,~L.; {\v{S}}mejkal,~L.; Strocov,~V.~N.; Constantinou,~P.; Hellenes,~A.~B.; Jaeschke~Ubiergo,~R.; Campos,~W.~H.; Bharadwaj,~V.~K.; Chakraborty,~A.; others Direct observation of altermagnetic band splitting in CrSb thin films. \emph{Nat. Commun.} \textbf{2024}, \emph{15}, 2116\relax
\mciteBstWouldAddEndPuncttrue
\mciteSetBstMidEndSepPunct{\mcitedefaultmidpunct}
{\mcitedefaultendpunct}{\mcitedefaultseppunct}\relax
\EndOfBibitem
\bibitem[Yang \latin{et~al.}(2025)Yang, Li, Yang, Li, Zheng, Zhu, Pan, Xu, Cao, Zhao, \latin{et~al.} others]{Yang-2025-CrSb}
Yang,~G.; Li,~Z.; Yang,~S.; Li,~J.; Zheng,~H.; Zhu,~W.; Pan,~Z.; Xu,~Y.; Cao,~S.; Zhao,~W.; others Three-dimensional mapping of the altermagnetic spin splitting in CrSb. \emph{Nat. Commun.} \textbf{2025}, \emph{16}, 1442\relax
\mciteBstWouldAddEndPuncttrue
\mciteSetBstMidEndSepPunct{\mcitedefaultmidpunct}
{\mcitedefaultendpunct}{\mcitedefaultseppunct}\relax
\EndOfBibitem
\bibitem[Zeng \latin{et~al.}(2024)Zeng, Zhu, Zhu, Liu, Ma, Hao, Liu, Qu, Yang, Jiang, \latin{et~al.} others]{Liuchang-2024-CrSb}
Zeng,~M.; Zhu,~M.-Y.; Zhu,~Y.-P.; Liu,~X.-R.; Ma,~X.-M.; Hao,~Y.-J.; Liu,~P.; Qu,~G.; Yang,~Y.; Jiang,~Z.; others Observation of Spin Splitting in Room-Temperature Metallic Antiferromagnet CrSb. \emph{Adv. Sci.} \textbf{2024}, \emph{11}, 2406529\relax
\mciteBstWouldAddEndPuncttrue
\mciteSetBstMidEndSepPunct{\mcitedefaultmidpunct}
{\mcitedefaultendpunct}{\mcitedefaultseppunct}\relax
\EndOfBibitem
\bibitem[Pan \latin{et~al.}(2024)Pan, Zhou, Lyu, Xiao, Yang, and Sun]{Pan-2024-stacking_alter}
Pan,~B.; Zhou,~P.; Lyu,~P.; Xiao,~H.; Yang,~X.; Sun,~L. General Stacking Theory for Altermagnetism in Bilayer Systems. \emph{Phys. Rev. Lett.} \textbf{2024}, \emph{133}, 166701\relax
\mciteBstWouldAddEndPuncttrue
\mciteSetBstMidEndSepPunct{\mcitedefaultmidpunct}
{\mcitedefaultendpunct}{\mcitedefaultseppunct}\relax
\EndOfBibitem
\bibitem[Sun \latin{et~al.}(2025)Sun, Wang, Yang, Huang, Ding, Dong, and Cheng]{Sun-2025-stacking_alter}
Sun,~W.; Wang,~W.; Yang,~C.; Huang,~S.; Ding,~N.; Dong,~S.; Cheng,~Z. Designing Spin Symmetry for Altermagnetism with Strong Magnetoelectric Coupling. \emph{Adv. Sci.} \textbf{2025}, e03235\relax
\mciteBstWouldAddEndPuncttrue
\mciteSetBstMidEndSepPunct{\mcitedefaultmidpunct}
{\mcitedefaultendpunct}{\mcitedefaultseppunct}\relax
\EndOfBibitem
\bibitem[Sun \latin{et~al.}(2025)Sun, Yang, Wang, Liu, Wang, Huang, and Cheng]{Sun-2025-alter}
Sun,~W.; Yang,~C.; Wang,~W.; Liu,~Y.; Wang,~X.; Huang,~S.; Cheng,~Z. Proposing Altermagnetic-Ferroelectric Type-III Multiferroics with Robust Magnetoelectric Coupling. \emph{Adv. Mater.} \textbf{2025}, \emph{37}, 2502575\relax
\mciteBstWouldAddEndPuncttrue
\mciteSetBstMidEndSepPunct{\mcitedefaultmidpunct}
{\mcitedefaultendpunct}{\mcitedefaultseppunct}\relax
\EndOfBibitem
\bibitem[Farooq \latin{et~al.}(2023)Farooq, Gui, and Huang]{2023umar}
Farooq,~M.~U.; Gui,~Z.; Huang,~L. Spontaneous spin momentum locking and anomalous Hall effect in ${\mathrm{BiFeO}}_{3}$. \emph{Phys. Rev. B} \textbf{2023}, \emph{107}, 075202\relax
\mciteBstWouldAddEndPuncttrue
\mciteSetBstMidEndSepPunct{\mcitedefaultmidpunct}
{\mcitedefaultendpunct}{\mcitedefaultseppunct}\relax
\EndOfBibitem
\bibitem[Gu \latin{et~al.}(2025)Gu, Liu, Zhu, Yananose, Chen, Hu, Stroppa, and Liu]{LiuQihang-2025-FEalter}
Gu,~M.; Liu,~Y.; Zhu,~H.; Yananose,~K.; Chen,~X.; Hu,~Y.; Stroppa,~A.; Liu,~Q. Ferroelectric Switchable Altermagnetism. \emph{Phys. Rev. Lett.} \textbf{2025}, \emph{134}, 106802\relax
\mciteBstWouldAddEndPuncttrue
\mciteSetBstMidEndSepPunct{\mcitedefaultmidpunct}
{\mcitedefaultendpunct}{\mcitedefaultseppunct}\relax
\EndOfBibitem
\bibitem[Duan \latin{et~al.}(2025)Duan, Zhang, Zhu, Liu, Zhang, \ifmmode \check{Z}\else \v{Z}\fi{}uti\ifmmode~\acute{c}\else \'{c}\fi{}, and Zhou]{DuanXuankai-2025-AFEalter}
Duan,~X.; Zhang,~J.; Zhu,~Z.; Liu,~Y.; Zhang,~Z.; \ifmmode \check{Z}\else \v{Z}\fi{}uti\ifmmode~\acute{c}\else \'{c}\fi{},~I.; Zhou,~T. Antiferroelectric Altermagnets: Antiferroelectricity Alters Magnets. \emph{Phys. Rev. Lett.} \textbf{2025}, \emph{134}, 106801\relax
\mciteBstWouldAddEndPuncttrue
\mciteSetBstMidEndSepPunct{\mcitedefaultmidpunct}
{\mcitedefaultendpunct}{\mcitedefaultseppunct}\relax
\EndOfBibitem
\bibitem[Li \latin{et~al.}(2013)Li, Cao, Niu, Shi, and Feng]{FengJi-2013-MnPX3-Valley-PNAS}
Li,~X.; Cao,~T.; Niu,~Q.; Shi,~J.; Feng,~J. Coupling the valley degree of freedom to antiferromagnetic order. \emph{Proc. Natl. Acad. Sci. U. S. A.} \textbf{2013}, \emph{110}, 3738--3742\relax
\mciteBstWouldAddEndPuncttrue
\mciteSetBstMidEndSepPunct{\mcitedefaultmidpunct}
{\mcitedefaultendpunct}{\mcitedefaultseppunct}\relax
\EndOfBibitem
\bibitem[Xu \latin{et~al.}(2023)Xu, Liu, Dai, Huang, and Wei]{WeiWei-2023-Mn2P2X3Y3-APL}
Xu,~Y.; Liu,~H.; Dai,~Y.; Huang,~B.; Wei,~W. Spin--valley splitting and spontaneous valley polarization in antiferromagnetic {Mn$_2$P$_2$X$_3$Y$_3$} monolayers. \emph{Appl. Phys. Lett.} \textbf{2023}, \emph{122}\relax
\mciteBstWouldAddEndPuncttrue
\mciteSetBstMidEndSepPunct{\mcitedefaultmidpunct}
{\mcitedefaultendpunct}{\mcitedefaultseppunct}\relax
\EndOfBibitem
\bibitem[{Fabian} \latin{et~al.}(2007){Fabian}, {Matos-Abiague}, {Ertler}, {Stano}, and {{\v{Z}}uti{\'c}}]{2007Igor}
{Fabian},~J.; {Matos-Abiague},~A.; {Ertler},~C.; {Stano},~P.; {{\v{Z}}uti{\'c}},~I. {Semiconductor spintronics}. \emph{Acta Phys. Slovaca} \textbf{2007}, \emph{57}, 565--907\relax
\mciteBstWouldAddEndPuncttrue
\mciteSetBstMidEndSepPunct{\mcitedefaultmidpunct}
{\mcitedefaultendpunct}{\mcitedefaultseppunct}\relax
\EndOfBibitem
\bibitem[Bl{\"o}chl(1994)]{Blochl-1994-PAW}
Bl{\"o}chl,~P.~E. Projector augmented-wave method. \emph{Phys. Rev. B} \textbf{1994}, \emph{50}, 17953\relax
\mciteBstWouldAddEndPuncttrue
\mciteSetBstMidEndSepPunct{\mcitedefaultmidpunct}
{\mcitedefaultendpunct}{\mcitedefaultseppunct}\relax
\EndOfBibitem
\bibitem[Kresse \latin{et~al.}(1994)Kresse, Furthm{\"u}ller, and Hafner]{kresse-1994-GGA}
Kresse,~G.; Furthm{\"u}ller,~J.; Hafner,~J. Theory of the crystal structures of selenium and tellurium: the effect of generalized-gradient corrections to the local-density approximation. \emph{Phys. Rev. B} \textbf{1994}, \emph{50}, 13181\relax
\mciteBstWouldAddEndPuncttrue
\mciteSetBstMidEndSepPunct{\mcitedefaultmidpunct}
{\mcitedefaultendpunct}{\mcitedefaultseppunct}\relax
\EndOfBibitem
\bibitem[Kresse and Furthm{\"u}ller(1996)Kresse, and Furthm{\"u}ller]{kresse-1996-efficient-iterative-schemes}
Kresse,~G.; Furthm{\"u}ller,~J. Efficient iterative schemes for ab initio total-energy calculations using a plane-wave basis set. \emph{Phys. Rev. B} \textbf{1996}, \emph{54}, 11169\relax
\mciteBstWouldAddEndPuncttrue
\mciteSetBstMidEndSepPunct{\mcitedefaultmidpunct}
{\mcitedefaultendpunct}{\mcitedefaultseppunct}\relax
\EndOfBibitem
\bibitem[Perdew \latin{et~al.}(1996)Perdew, Burke, and Ernzerhof]{perdew-1996-PBE}
Perdew,~J.~P.; Burke,~K.; Ernzerhof,~M. Generalized gradient approximation made simple. \emph{Phys. Rev. Lett.} \textbf{1996}, \emph{77}, 3865\relax
\mciteBstWouldAddEndPuncttrue
\mciteSetBstMidEndSepPunct{\mcitedefaultmidpunct}
{\mcitedefaultendpunct}{\mcitedefaultseppunct}\relax
\EndOfBibitem
\bibitem[Kresse and Joubert(1999)Kresse, and Joubert]{kresse-1999-PAW}
Kresse,~G.; Joubert,~D. From ultrasoft pseudopotentials to the projector augmented-wave method. \emph{Phys. Rev. B} \textbf{1999}, \emph{59}, 1758\relax
\mciteBstWouldAddEndPuncttrue
\mciteSetBstMidEndSepPunct{\mcitedefaultmidpunct}
{\mcitedefaultendpunct}{\mcitedefaultseppunct}\relax
\EndOfBibitem
\bibitem[Chittari \latin{et~al.}(2016)Chittari, Park, Lee, Han, MacDonald, Hwang, and Jung]{MacDonald-2016-MPX3-PRB}
Chittari,~B.~L.; Park,~Y.; Lee,~D.; Han,~M.; MacDonald,~A.~H.; Hwang,~E.; Jung,~J. Electronic and magnetic properties of single-layer MPX 3 metal phosphorous trichalcogenides. \emph{Phys. Rev. B} \textbf{2016}, \emph{94}, 184428\relax
\mciteBstWouldAddEndPuncttrue
\mciteSetBstMidEndSepPunct{\mcitedefaultmidpunct}
{\mcitedefaultendpunct}{\mcitedefaultseppunct}\relax
\EndOfBibitem
\bibitem[Zhou \latin{et~al.}(2023)Zhou, Zheng, Li, Zhang, and Ouyang]{Fangping-2023-MnPTe3-PRB}
Zhou,~W.; Zheng,~G.; Li,~A.; Zhang,~D.; Ouyang,~F. Orbital contribution to the regulation of the spin-valley coupling in antiferromagnetic monolayer {MnPTe$_3$}. \emph{Phys. Rev. B} \textbf{2023}, \emph{107}, 035139\relax
\mciteBstWouldAddEndPuncttrue
\mciteSetBstMidEndSepPunct{\mcitedefaultmidpunct}
{\mcitedefaultendpunct}{\mcitedefaultseppunct}\relax
\EndOfBibitem
\bibitem[Huang \latin{et~al.}(2018)Huang, Du, Wu, Xiang, Deng, and Kan]{huang-2018-CrX3prediction-PRL}
Huang,~C.; Du,~Y.; Wu,~H.; Xiang,~H.; Deng,~K.; Kan,~E. Prediction of intrinsic ferromagnetic ferroelectricity in a transition-metal halide monolayer. \emph{Phys. Rev. Lett.} \textbf{2018}, \emph{120}, 147601\relax
\mciteBstWouldAddEndPuncttrue
\mciteSetBstMidEndSepPunct{\mcitedefaultmidpunct}
{\mcitedefaultendpunct}{\mcitedefaultseppunct}\relax
\EndOfBibitem
\bibitem[Xue \latin{et~al.}(2019)Xue, Hou, Wang, and Wu]{xue-2019-CrCl3-PRB}
Xue,~F.; Hou,~Y.; Wang,~Z.; Wu,~R. Two-dimensional ferromagnetic van der Waals CrCl$_3$ monolayer with enhanced anisotropy and Curie temperature. \emph{Phys. Rev. B} \textbf{2019}, \emph{100}, 224429\relax
\mciteBstWouldAddEndPuncttrue
\mciteSetBstMidEndSepPunct{\mcitedefaultmidpunct}
{\mcitedefaultendpunct}{\mcitedefaultseppunct}\relax
\EndOfBibitem
\bibitem[Grimme \latin{et~al.}(2011)Grimme, Ehrlich, and Goerigk]{Grimme-2011-DFT_D3-JCC}
Grimme,~S.; Ehrlich,~S.; Goerigk,~L. Effect of the damping function in dispersion corrected density functional theory. \emph{J. Comput. Chem.} \textbf{2011}, \emph{32}, 1456--1465\relax
\mciteBstWouldAddEndPuncttrue
\mciteSetBstMidEndSepPunct{\mcitedefaultmidpunct}
{\mcitedefaultendpunct}{\mcitedefaultseppunct}\relax
\EndOfBibitem
\bibitem[Kim \latin{et~al.}(2022)Kim, Kim, Cheon, and Kim]{Kim-2022-VASPBERRY}
Kim,~S.-W.; Kim,~H.-J.; Cheon,~S.; Kim,~T.-H. Circular Dichroism of Emergent Chiral Stacking Orders in Quasi-One-Dimensional Charge Density Waves. \emph{Phys. Rev. Lett.} \textbf{2022}, \emph{128}, 046401\relax
\mciteBstWouldAddEndPuncttrue
\mciteSetBstMidEndSepPunct{\mcitedefaultmidpunct}
{\mcitedefaultendpunct}{\mcitedefaultseppunct}\relax
\EndOfBibitem
\bibitem[Wu \latin{et~al.}(2018)Wu, Zhang, Song, Troyer, and Soluyanov]{WannierTools}
Wu,~Q.; Zhang,~S.; Song,~H.-F.; Troyer,~M.; Soluyanov,~A.~A. WannierTools : An open-source software package for novel topological materials. \emph{Comput. Phys. Commun.} \textbf{2018}, \emph{224}, 405 -- 416\relax
\mciteBstWouldAddEndPuncttrue
\mciteSetBstMidEndSepPunct{\mcitedefaultmidpunct}
{\mcitedefaultendpunct}{\mcitedefaultseppunct}\relax
\EndOfBibitem
\bibitem[Mostofi \latin{et~al.}(2014)Mostofi, Yates, Pizzi, Lee, Souza, Vanderbilt, and Marzari]{Wannier90}
Mostofi,~A.~A.; Yates,~J.~R.; Pizzi,~G.; Lee,~Y.-S.; Souza,~I.; Vanderbilt,~D.; Marzari,~N. An updated version of wannier90: A tool for obtaining maximally-localised Wannier functions. \emph{Comput. Phys. Commun.} \textbf{2014}, \emph{185}, 2309 -- 2310\relax
\mciteBstWouldAddEndPuncttrue
\mciteSetBstMidEndSepPunct{\mcitedefaultmidpunct}
{\mcitedefaultendpunct}{\mcitedefaultseppunct}\relax
\EndOfBibitem
\end{mcitethebibliography}

\end{document}


\newpage

\subsection{Model $k \cdot p $ Halmiltonians of Altermagnets}

In our analysis, the mirror operation \(\mathcal{M}_y\) is defined as a reflection with the normal vector along the \(y\)-axis. Under this symmetry operation, the spatial coordinates transform as
\begin{equation}
    \mathcal{M}_y: (x, y, z) \rightarrow (x, -y, z).
\end{equation}

Accordingly, the spin and momentum components transform as follows:
\begin{align}
    &\mathcal{M}_y S_x \rightarrow -S_x, \quad \mathcal{M}_y S_y \rightarrow S_y, \quad \mathcal{M}_y S_z \rightarrow S_z, \\
    &\mathcal{M}_y k_x \rightarrow k_x, \quad \mathcal{M}_y k_y \rightarrow -k_y, \quad \mathcal{M}_y k_z \rightarrow k_z.
\end{align}

As a result, the spin textures in momentum space transform under \(\mathcal{M}_y\) as:
\begin{align}
    &\mathcal{M}_y: S_x(k_x, k_y) \rightarrow -S_x(k_x, -k_y), \\
    &\mathcal{M}_y: S_y(k_x, k_y) \rightarrow \phantom{-}S_y(k_x, -k_y), \\
    &\mathcal{M}_y: S_z(k_x, k_y) \rightarrow -S_z(k_x, -k_y).
\end{align}

Therefore, both \(S_x(k_x, k_y)\) and \(S_z(k_x, k_y)\) are odd functions with respect to \(k_y\). Retaining only the leading-order terms in a momentum expansion, one obtains
\begin{equation}
    S_x = \delta_x k_y, \quad S_z = \delta_z k_y,
\end{equation}
which should vanish at K point and only the components linear in \(k_y\) are kept for simplicity.

Focusing on the spin polarization along the \(z\)-axis, this leads to a symmetry-allowed term of the form \(\delta_z k_y \sigma_z\), which characterizes the spin texture of altermagnetic states. In the vicinity of the valley, this contribution can be compactly written as
\begin{equation}
    \delta \, \sigma_0 \tau_0 S_z k_y,
\end{equation}
where \(\delta\) is a constant, and \(\sigma_0\), \(\tau_0\) are identity matrices as defined earlier.

\begin{figure*}
  \centering
  \includegraphics[scale=0.55]{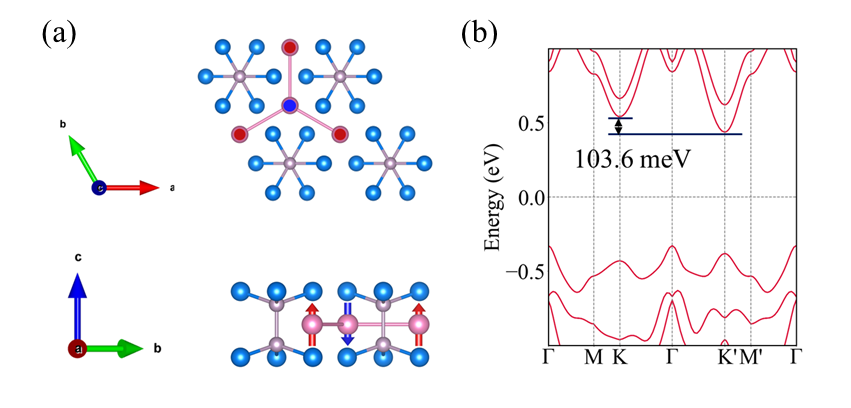}
  \caption{
  Crystal structure and band structure (with SOC) of monolayer MnPTe$_3$.}
  \label{BC_vasp}
\end{figure*}

\begin{figure*}
  \centering
  \includegraphics[scale=0.5]{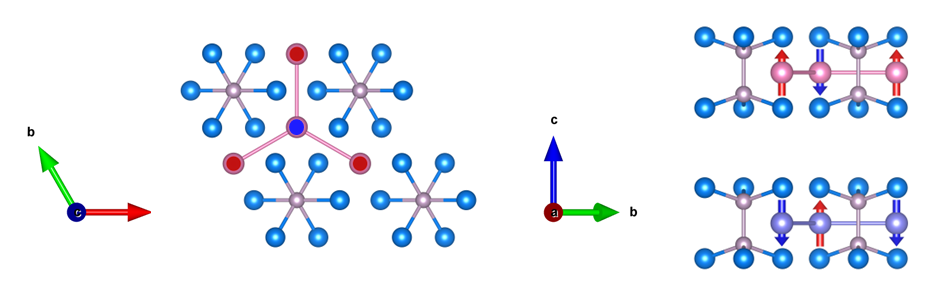}
  \caption{
  Crystal structure of AA-stacking order of bilayer MnPTe$_3$.}
  \label{BC_vasp}
\end{figure*}

\begin{table*}
\caption{\label{tab:table1} The parameters (in eV) of the \(k\cdot p\) model of bilayer MnPTe$_3$ obtained by fitting the DFT band structures.}
\centering
\begin{tabular}{lccccccc}
\hline
 &$v_f$ & $m$ & $\lambda_c$ & $\lambda_c'$ & $\lambda_v$ & $\lambda_v'$  \\
\hline
AB-stacking & 2.16 & 0.4210 & 0.0408 & 0.0650&0.0100&0.0073 \\
BA-stacking & 2.16 & 0.4108& 0.0680 & 0.0430 & 0.0080 & 0.0100  \\
\hline
\end{tabular}

\begin{tabular}{cccccc}
\hline
  $\phi$ & $\phi'$ & $\delta$ & $U_E$  \\
\hline
 0.0060 & 0.0060 & -0.0100 & 0.0400 \\
 -0.0060  & -0.0060 & -0.0100 & -0.0400  \\
\hline
\end{tabular}
\end{table*}

\begin{table*}
\caption{\label{tab:table1} Berry curvature (in \AA$^2$) at the CBM of bilayer MnPTe$_3$, obtained by fitting the DFT band structure using the \(k \cdot p\) model.}
\centering
\begin{tabular}{lcc}
\hline
 & $\Omega_K$ & $\Omega_{K'}$ \\
\hline
AB-stacking & -17.9402 & -14.0242 \\
BA-stacking & 14.9514 & 19.0432 \\
\hline
\end{tabular}
\end{table*}

\begin{figure*}
  \centering
  \includegraphics[scale=0.5]{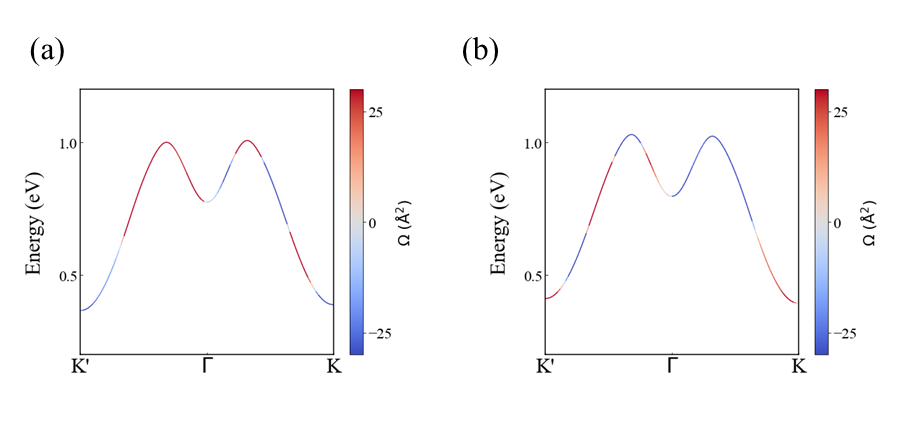}
  \caption{
  Berry curvature at the CBM for AB- and BA-stacked bilayer MnPTe$_3$, as calculated using VASPBERRY, is shown in (a) and (b).}
  \label{BC_vasp}
\end{figure*}

\begin{figure*}
  \centering
  \includegraphics[scale=0.6]{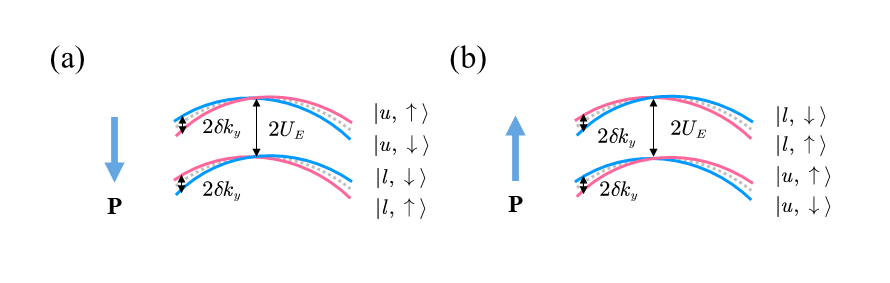}
  \caption{
  Nonrelativistic spin splittings of the top valence bands of bilayer MnPTe$_3$ along \(\Gamma - K\) path from the proposed model in the main text. The spin splitting pattern away from K point is consistent with our DFT calculations.}
  \label{BC_vasp}
\end{figure*}

\begin{figure*}
  \centering
  \includegraphics[scale=0.6]{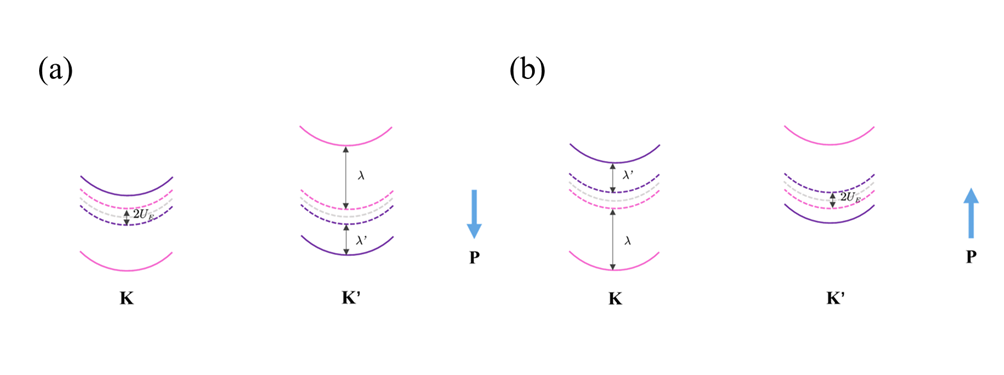}
  \caption{The grey dashed line represents the case without electrostatic potential and SOC. The pink and purple dashed lines indicate the conduction bands at the \(K\) and \(K'\) points in the upper and lower layers, respectively, under electrostatic potential but without SOC. The corresponding solid lines represent the conduction bands when both electrostatic potential and SOC are included.  }
  \label{BC_vasp}
\end{figure*}

\begin{figure*}
  \centering
  \includegraphics[scale=0.5]{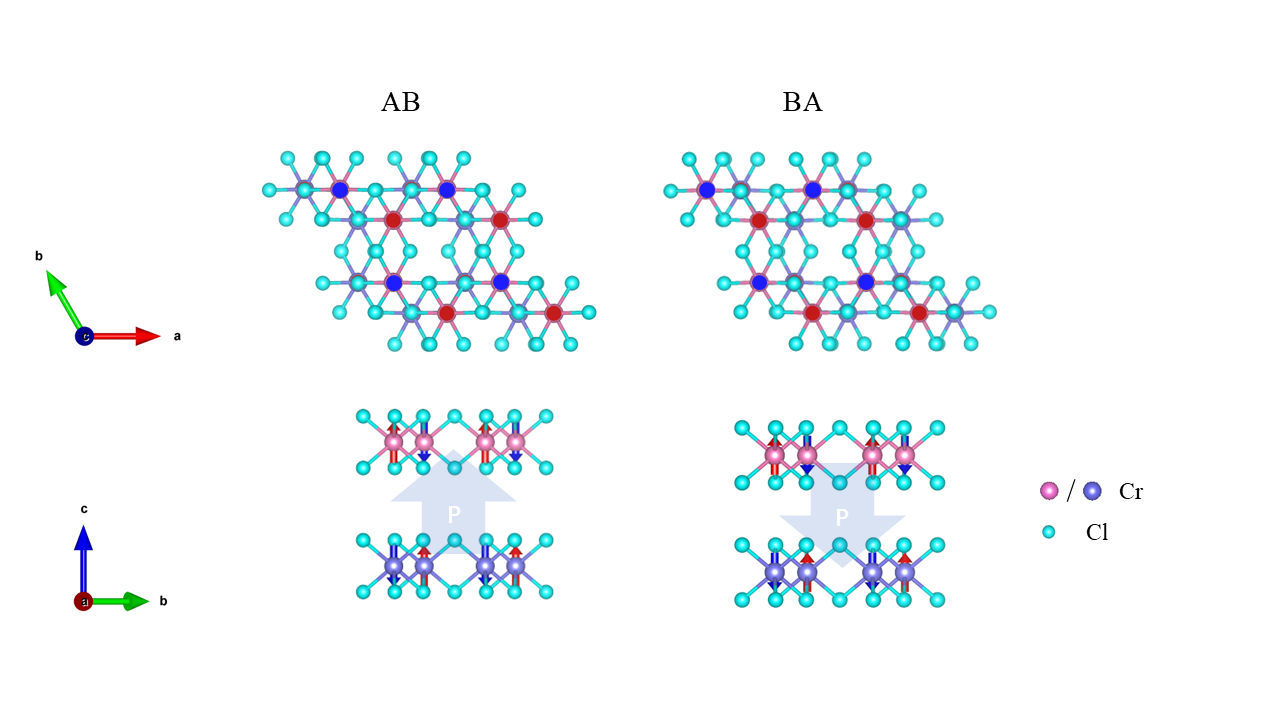}
  \caption{Crystal structures of AB- and BA-stacked bilayer CrCl$_3$ under -6\% biaxial strain, which transforms the ground state of monolayer CrCl$_3$ into an intralayer antiferromagnetic configuration.}
  \label{BC_vasp}
\end{figure*}

\begin{figure*}
\includegraphics[scale=0.5]{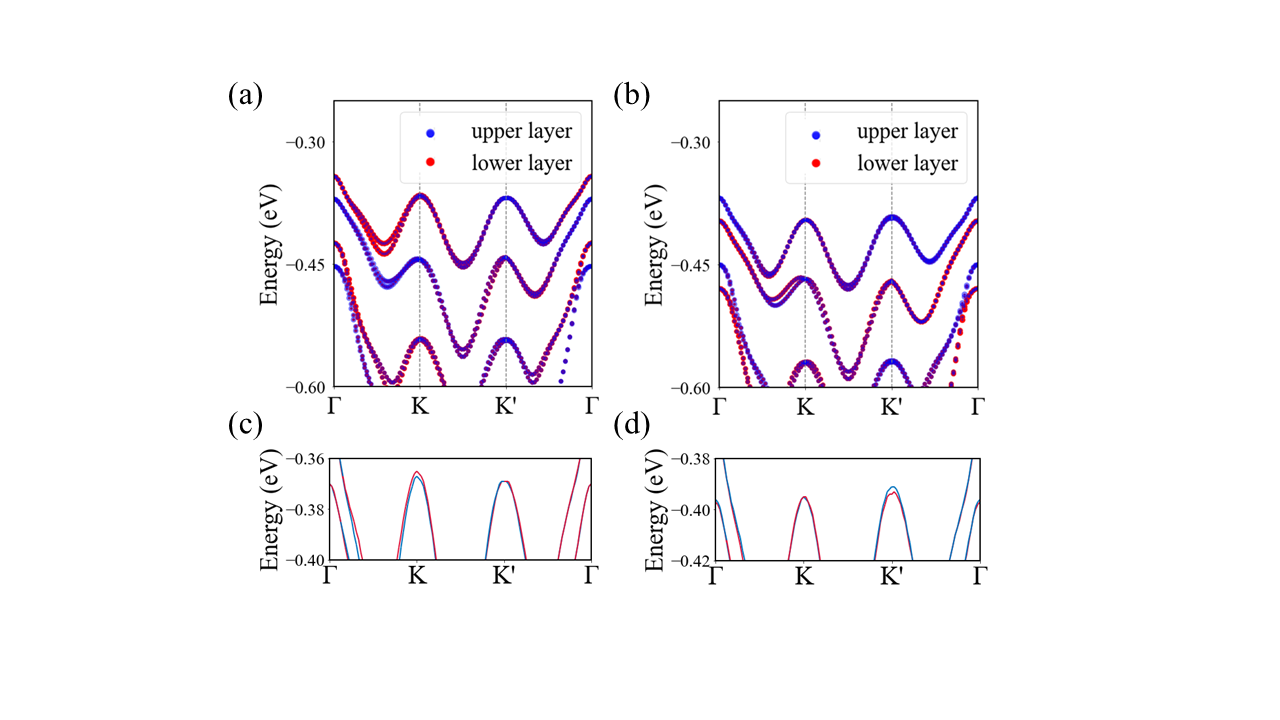}
\caption{(a)-(b) Projected band structures at the VBM of AB- and BA-stacked bilayer CrCl$_3$ with SOC included. Blue and red dots represent the contributions from the upper and lower layers, respectively. CrCl$_3$ serves as an example with weak SOC, where the valleys in the two layers do not exhibit nesting. Under ferroelectric polarization reversal, the exchange of layer contributions leads to a 4 meV valley polarization reversal. (c)-(d) Magnified views of the conduction band edges in (a) and (b), showing the spin projections along the $z$-axis. Red and blue colors denote positive and negative spin-$z$ components, respectively.
}\label{band_soc}
\end{figure*}

\begin{figure*}
  \centering
  \includegraphics[scale=0.5]{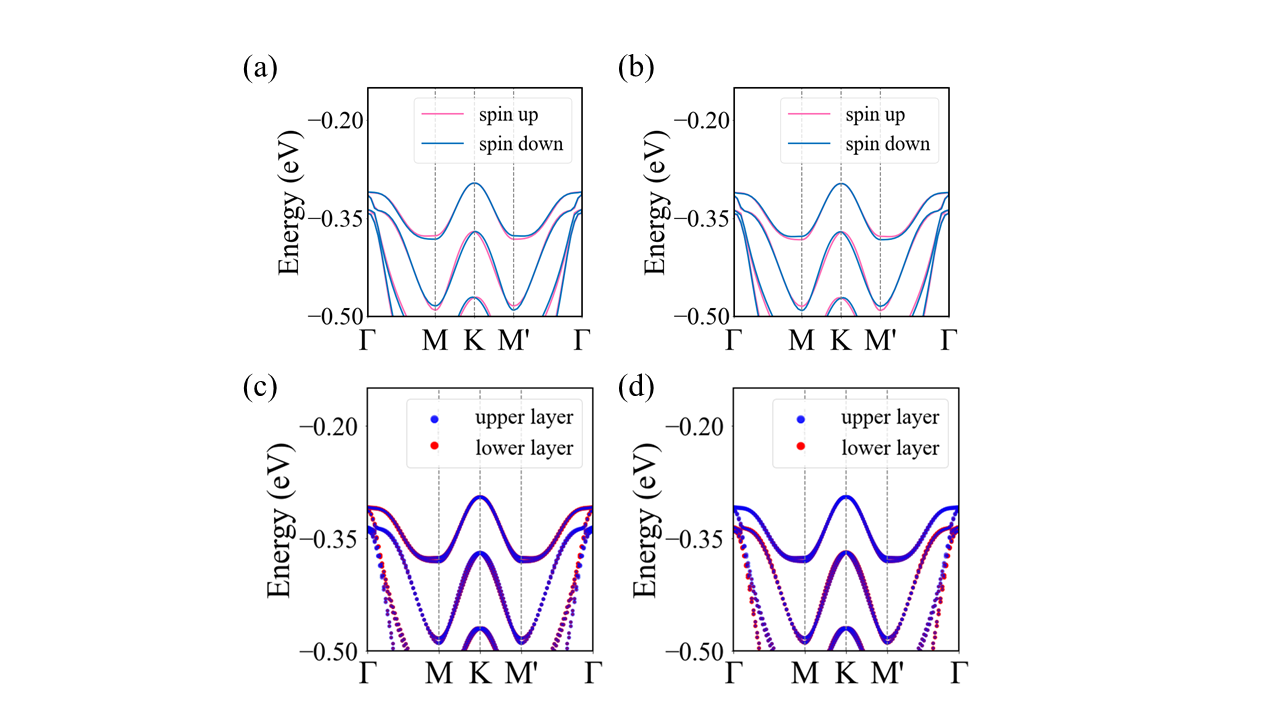}
  \caption{(a)–(b) Band structures at the VBM of AB- and BA-stacked bilayer CrCl\textsubscript{3} without SOC. (c)–(d) Projected band structures at the VBM of AB- and BA-stacked bilayer CrCl\textsubscript{3} without SOC, where red and blue dots indicate contributions from the lower and upper layers, respectively.}
  \label{BC_vasp}
\end{figure*}